\documentclass[draft]{agujournal2019}
\usepackage{url} 
\usepackage{lineno}
\usepackage[inline]{trackchanges} 
\usepackage{soul}
\usepackage{graphicx}

\draftfalse

\journalname{JGR: Space Physics}

\begin{document}


\title{Determining the beaming of Io decametric emissions : a remote diagnostic to probe the Io-Jupiter interaction}

\authors{L. Lamy\affil{1,2,3}, L. Colomban\affil{4}, P. Zarka\affil{1,2}, R. Prang\'e\affil{1}, M. S. Marques\affil{5},  C. Louis\affil{6}, W. Kurth\affil{7}, B. Cecconi\affil{1,2}, J. Girard\affil{1,2}, J.-M. Griessmeier\affil{2,4}, S. Yerin\affil{8,9}}

\affiliation{1}{LESIA, Observatoire de Paris, Universit\'e PSL, CNRS, Sorbonne Universit\'e, Universit\'e de Paris, Meudon, France}
\affiliation{2}{Station de Radioastronomie de Nan\c cay, Observatoire de Paris, Universit\'e PSL, CNRS, Univ. Orl\'eans, Nan\c cay, France}
\affiliation{3}{LAM, Pyth\'eas, Aix Marseille Universit\'e, CNRS, CNES, Marseille, France}
\affiliation{4}{LPC2E, CNRS, Universit\'e d'Orl\'eans, 3A avenue de la Recherche Scientifique, Orl\'eans, France}
\affiliation{5}{Departamento de Geofisica, Universidade Federal do Rio Grande do Norte, Natal, Brazil}
\affiliation{6}{DIAS, Dublin, Ireland}
\affiliation{7}{Department of Physics and Astronomy, University of Iowa, Iowa City, Iowa, USA}
\affiliation{8}{Institute of Radio Astronomy of NAS of Ukraine, Kharkiv, Ukraine}
\affiliation{9}{V. N. Karazin Kharkiv National University, Kharkiv, Ukraine}

\correspondingauthor{Laurent Lamy}{laurent.lamy@obspm.fr}

\begin{keypoints}

\item We derive the Io-decametric emission angle $\theta$ from Juno, Nan\c cay Decameter Array and NenuFAR data using 3 methods to locate the radiosources


\item $\theta(f)$ decreases from $75^\circ-80^\circ$ to $70^\circ-75^\circ$ over the range $10-40$~MHz and varies as a function of time (or longitude of Io)


\item The inferred electron energies amplifying Io-decametric waves range from 3 to 16 keV also vary as a function of altitude and time

\end{keypoints}

%
%

%
%


\begin{abstract}
We investigate the beaming of 11 Io-Jupiter decametric (Io-DAM) emissions observed by Juno/Waves, the Nan\c cay Decameter Array and NenuFAR. Using an up-to-date magnetic field model and three methods to position the active Io Flux Tube (IFT), we accurately locate the radiosources and determine their emission angle $\theta$ from the local magnetic field vector. These methods use (i) updated models of the IFT equatorial lead angle, (ii) ultraviolet (UV) images of Jupiter's aurorae and (iii) multi-point radio measurements. The kinetic energy $E_{e-}$ of source electrons is then inferred from $\theta$ in the framework of the Cyclotron Maser Instability. The precise position of the active IFT achieved from methods (ii,iii) can be used to test the effective torus plasma density. Simultaneous radio/UV observations reveal that multiple Io-DAM arcs are associated with multiple UV spots and provide the first direct evidence of an Io-DAM arc associated with a trans-hemispheric beam UV spot. Multi-point radio observations probe the Io-DAM sources at various altitudes, times and hemispheres. Overall, $\theta$ varies a function of frequency (altitude), by decreasing from $75^\circ-80^\circ$ to $70^\circ-75^\circ$ over $10-40$ MHz with slightly larger values in the northern hemisphere, and independently varies as a function of time (or longitude of Io). Its uncertainty of a few degrees is dominated by the error on the longitude of the active IFT. The inferred values of $E_{e-}$ also vary as a function of altitude and time. For the 11 investigated cases, they range from 3 to 16 keV, with a $6.6\pm2.7$ keV average.
\end{abstract}

\section*{Plain Language Summary}
The auroral decametric emissions of Jupiter induced by Io (Io-DAM) are radiated along high latitude magnetic field lines at large aperture angles from the local magnetic field vector, forming a thin hollow cone. In this study, we determine the emission angle $\theta$ of 11 cases of Io-DAM emissions observed by Juno/Waves, the Nancay Decameter Array and the NenuFAR radiotelescope with an up-to-date magnetic field model and three different methods aimed at minimizing the uncertainty on $\theta$. These methods accurately position the active Io magnetic Flux Tube (IFT) which hosts the decametric radiosources by using (i) models of the active IFT, (ii) ultraviolet images of Jupiter's aurorae and (iii) multi-point radio measurements. Most notably, we found that $\theta$ varies within $70^\circ-80^\circ$ as a function of the source altitude along the field line and independently as a function of time. Assuming that the Io-DAM emissions are driven by the Cyclotron Maser Instability from energetic electrons, we infer from the measured $\theta$ the kinetic energy $E_{e-}$ of the source electrons accelerated by the Io-Jupiter interaction. The obtained values of $E_{e-}$ also depend on altitude and time and vary between 3 and 16~keV, with a $\sim$6.5~keV average, in agreement with Juno in situ measurements.

\section{Introduction}
 
The motion of Io through the intense magnetic field prevailing in the inner Jovian magnetosphere sustains an Alfv\'enic electric current system \cite[and refs therein]{Saur_JGR_04,Hess_JGR_11}. This current system mainly develops along the so-called active Io Flux Tube (IFT) which leads in longitude (in the System III system) the instantaneous IFT by a time-variable equatorial lead angle $\delta$, resulting from the finite time needed by the Alfv\'enic perturbations carrying the current to exit the dense Io plasma torus. The transmission and reflection of Alfv\'enic perturbations on the edges of the Io plasma torus and/or the Jovian ionosphere additionally transfer part of the total current along secondary flux tubes. Overall, this current system can accelerate electrons which in turn drive powerful radio emissions at decametric (DAM) wavelengths, up to 40~MHz, above the ionosphere and bright ultraviolet (UV) footprints in the atmosphere \cite[and refs therein]{Prange_Nature_96,Clarke_Science_96,Clarke_book_04,Hess_JGR_10,Badman_SSR_15}. The characteristics of Jovian auroral emissions induced by Io, which can be monitored remotely by Earth-based telescopes or in situ by exploration spacecraft, can in turn be employed to probe the Io-Jupiter electrodynamic interaction. For instance, the multiplicity of Io UV spots, made of a Main Alfv\'en Wing (MAW) spot and a series of Reflected Alfv\'en Wing (RAW) ones, maps the instantaneous topology of the Alfv\'enic current system within and outside the Io torus \cite{Bonfond_GRL_08}. 


Jupiter's auroral radio emissions have been regularly observed above the terrestrial ionospheric cutoff of 10~MHz from the ground since the 1950s \cite{Burke_JGR_55}, which incidentally yielded the discovery of a Jovian magnetic field and the first estimates of its surface magnitude. A variety of ground-based radiotelescopes have been successively involved in the long-term monitoring of Jovian DAM emissions, from the historical Clarke Lake decametric array up to facilities still in operation such as the Nan\c cay Decameter Array (NDA) or under construction like NenuFAR. The DAM component was early found to be dominated by emissions induced (or controlled) by Io (hereafter Io-DAM) \cite{Bigg_Nature_64}. The low frequency part of Jovian auroral radio emissions, ranging from hectometric (HOM) emissions at a few MHz down to kilometric (KOM) emissions at a few 10-100~kHz, has been observed by exploration probes such as Voyager 1 \& 2, Ulysses, Cassini, Galileo up to Juno. Thanks to this large observational dataset, the macroscopic properties of DAM/HOM/KOM components are now well established. For our purpose, we will simply remind that these radio waves are produced dominantly on the eXtraodinary (X) free-space mode above the ionosphere at frequencies $f$ near the local electron gyrofrequency $f_{ce}$ (proportional to the local magnetic field magnitude $B$), that they are strongly circularly polarized with a Right-Handed (RH) or Left-Handed (LH) sense for northern and southern emissions, respectively, and that the amplified waves are radiated along a thin hollow cone (a few degrees wide) at large aperture angles from the local magnetic field vector. The similarity of Jovian auroral radio emissions with the Terrestrial Kilometric Radiation early led to the postulate of a common generation mechanism, namely the Cyclotron Maser Instability (CMI) which amplifies radio waves in strongly magnetized and depleted regions from mildly relativistic non-maxwellian electrons \cite[and refs therein]{Zarka_JGR_98, Treumann_AAR_06}. Recent in situ measurements of the Juno spacecraft within the Jovian auroral regions confirmed that non-Io auroral radio emissions are indeed driven by the CMI \cite{Louarn_GRL_17}. 

Io-DAM emissions distinguish from non-Io DAM ones by being more intense and reaching higher frequencies (the latter being confined below 27~MHz). They also display characteristic arc-shaped structures in the time-frequency (t-f) plane, and have been historically classified in four main categories (termed A, B, C and D) depending on their sense of curvature and hemisphere of origin \cite[and refs therein]{Carr_83,Marques_AA_17}. In brief, A and B (C and D, resp.) arcs are RH (LH) polarized and correspond to northern (southern, resp.) extraordinary mode emissions. A and C (B and D, resp.) arcs display a vertex-late/closed parenthesis (vertex-early/open parenthesis) shape corresponding to an eastward (westward) position of the emitting flux tube with respect to the observer. This arc-shaped topology directly results from the anisotropic emission beaming pattern coupled to the motion of the 'radio-active' magnetic flux tube with respect to the observer. Assessing the beaming pattern of Io-DAM arcs has been a continuous matter of interest in the CMI framework \cite[and refs therein]{Ray_JGR_08}. Assuming straight line propagation, a central issue in the determination of the Io-DAM frequency-dependent emission angle relies on the correct positioning of the active IFT hosting the radiosources. This requires an accurate knowledge of the magnetic field model - a variety of which has been used in the literature - and of the lead angle - which was not always taken into account to determine empirical beaming angle models. Attempts to retrieve the beaming pattern from ray tracing were even more uncertain, as they additionally require a realistic plasma model.

Extending the work of \citeA{Queinnec_JGR_98}, \citeA{Hess_GRL_08} successfully reproduced the characteristic t-f shape of Io-DAM arcs thanks the geometrical simulation code ExPRES (Exoplanetary and Planetary Radio Emission Simulator). The code computes the visibility of radiosources spread in frequency (altitude) along a chosen flux tube for a given observer. The waves propagate in straight line with an initial aperture angle $\theta(f) = ({\bf k}, {\bf B})$ ({\bf k} wave vector at the source, {\bf B} local magnetic field vector) that is theoretically computed in the CMI framework (the interested reader is referred to an exhaustive presentation of ExPRES in \cite{Louis_AA_19}). The Io-DAM arcs could only be reproduced for emission oblique with respect to {\bf B}, with $\theta(f)$ decreasing with frequency. The authors simulated it from loss cone-driven CMI through a simple formalism which links $\theta(f)$ to the velocity of CMI-resonant electrons $v$ accelerated by the Io-Jupiter current system (see below). For the 6 cases of Io-DAM arcs investigated by \citeA{Hess_GRL_08} with the VIT4 magnetic field model, best fits were obtained for $\theta$ decreasing from $\sim80^\circ$ to $\sim40^\circ$, with $\delta$ in the $6-30^\circ$ range and inferred CMI-unstable electron kinetic energies $E_{e-}=3$~keV in the south and 0.64 keV in the north. In a follow-up study, \citeA{Hess_PSS_10} used the VIP4 magnetic field model to fit the shape of 50 southern Io-DAM arcs and derived $\delta$ ranging from $0$ to $10^\circ$ and $E_{e-}$ of a few keV sometimes as high as 20~keV. Both quantities were noticed to vary with Io's System III (SIII) longitude. The main limitations of those pre-Juno studies come from the poor knowledge of the magnetic field, which in turn yields large uncertainties in the location of the radiosources, in the orientation of $\bf{B}$ at the source and in subsequent models of the equatorial lead angle $\delta$ \cite{Hess_PRE8_17}.

The in situ polar exploration of Jupiter's auroral regions by Juno since mid-2016 therefore offered the possibility to re-assess the Io-DAM beaming angle with minimal uncertainties. Firstly, magnetic measurements acquired during the first nine polar orbits of Juno served to compute an updated magnetic field model up to order 10, labelled JRM09 \cite{Connerney_GRL_18}. In addition, the ongoing analysis of Juno encounters with the active IFT interestingly confirmed the ubiquity of CMI-unstable loss cone electron distribution functions with $E_{e-}$ ranging from 1 to 26~keV \cite{Louis_AGU_20}.

In this context, two independent studies based on JRM09 re-investigated the beaming of Io-DAM arcs while using the formalism of \cite{Hess_GRL_08} to assess the CMI-unstable electron energy. \citeA{Wang_EPP_20} developed a method based on multi-point radio observations of Io-DAM to constrain the locus of radiosources and to derive accurate beaming angles. Applied to Wind/STEREO observations of a single Io-DAM event, the authors obtained $\theta(f)$ varying within $\sim65-60^\circ$ for $\delta\sim32^\circ$, leading to $E_{e-}\sim10-20$~keV. \citeA{Martos_JGR_20} investigated 4 cases of Io-DAM events observed by Juno and jointly derived $\theta(f)$ and $\delta$. They obtained highly variable values within $\sim33-85^\circ$ and $\sim1-40^\circ$, respectively, leading to $E_{e-}$ ranging from 1 to 50~keV. 

The present study builds up on the work of \citeA{Queinnec_JGR_98,Hess_GRL_08} and aims at accurately determining the beaming of a series of Io-DAM emissions observed by Juno, the NDA and NenuFAR, using a Juno-derived magnetic field model (JRM09 and current sheet model). The novelty of our approach consists in the development of three different methods to accurately constrain the location of the active IFT. These use (i) updated models of the Io equatorial lead angle, (ii) UV images, here obtained with the Hubble Space Telescope (HST), simultaneous to the radio observations and (iii) multi-point radio measurements. Section \ref{data} presents the dataset and section \ref{method} introduces our methodology. Sections \ref{lead_angle} to \ref{radio_radio} utilize methods (i), (ii) and (iii) on case studies. Section \ref{stand_alone} prefigures the statistical studies that will be based on single point radio observations. Section \ref{discussion} discusses our results, while a summary of those and perspectives are dealt with in section \ref{summary}. 


\section{Dataset}
\label{data}

The observational dataset analyzed in this study corresponds to radio observations of Jupiter at decametric wavelengths acquired by the NASA Juno spacecraft, in orbit around Jupiter since mid-2016, by the Nan\c cay Decameter Array and the NenuFAR radiotelescopes, both located at the Nan\c cay radioastronomy station (Sologne, France), complemented by HST images of Jupiter's UV aurorae. 

 \subsection{Juno/Waves}

The Waves instrument onboard the Juno spacecraft measures electric fields sensed by a single dipole antenna with a 2.41~m effective length, connected to four receivers sampling frequencies ranging from 50~Hz up to 41~MHz \cite{Kurth_SSR_17}. Hereafter, we focus on wideband survey observations of the flux density (Juno/Waves does not provide polarization measurements), acquired at a cadence of 1 spectrum/s by the HFR-Hi(gh) swept-frequency receiver, covering the 3~MHz-41~MHz range with linearly-spaced channels every 1~MHz. Data from the HFR-Low receiver (covering the $100$~kHz$-3$~MHz range), less sensitive and subject to strong, time variable interferences, can be used only during source crossings.

\subsection{The Nan\c cay Decameter Array}

The Nan\c cay Decameter Array (NDA) is an historical radiotelescope of the Nan\c cay radioastronomy station operating in the 10-100~MHz range \cite{Boischot_Icarus_80,Lecacheux_00}. It is a phased arrray made of 144 helical "Tee-Pee" antennae, corresponding to a $\sim7000$~m$^2$ effective area at 25 MHz. The NDA is composed of two sub-arrays of 72 antennae each, sensitive to Right-Handed (RH) and Left-Handed (LH) polarization. The NDA has been observing Jupiter on a quasi-daily basis since January 1978 with receivers of increasingly improved performances, the latest of which was specifically developed to collect observations at very high t-f resolution in support of Juno \cite{Zarka_Goutelas_11,Lamy_PRE8_17}. Here, we focus on survey observations acquired at a 1~s temporal cadence by the swept-frequency Routine receiver over 10-40~MHz, with linearly-spaced frequency channels every 75~kHz.

\subsection{NenuFAR}

The NenuFAR telescope is a giant phased array and interferometer currently under construction in Nan\c cay \cite{Zarka_URSI_20}. As part of the Early Science phase which started mid-2019, NenuFAR regularly observes Jupiter in support of Juno through a dedicated Jupiter Key Project. In late 2019, when the data analyzed in this study were acquired, NenuFAR was composed of 56 mini-arrays, each made of 19 crossed-dipoles, reaching an effective area of $\sim31000$~m$^2$ at 27~MHz. Survey observations of Jupiter were obtained with the UnDySPuTeD receiver, which computes here the full Stokes parameters, over 10-40~MHz, with a 84~ms temporal cadence and linearly-spaced frequency channels every 12.2~kHz.

\subsection{HST images}

The Far-UV (FUV) images of the Jovian aurorae used in combination with the radio observations in section \ref{radio_uv} were acquired by the Space Telescope Imaging Spectrograph (STIS) onboard HST and retrieved from the APIS service \cite{Lamy_AC_15}. For our purpose, we used the STIS time-tag capability to produce 100~s sequenced images from which we built polar projections at the 900~km peak altitude for the MAW Io footprint, slightly higher than the 400~km altitude of the main auroral arc \cite{Bonfond_JGR_09}. 

\section{Methodology}
\label{method}

This section describes the methodology employed to determine the Io-DAM beaming andxt the energy of underlying CMI-driving source electrons from time-frequency observations of Io-DAM emissions.

\subsection{Determination of the Io-DAM beaming pattern}
\label{beaming}

To determine the opening angle of the emission cone at the source for a given Io-DAM arc, we proceed as follows. (1) We first fit the arc of interest in the dynamic spectrum by visually tracking intensity maxima with a set of t-f coordinates. For distant observers, the time is corrected for light time travel to be propagated back to Jupiter. (2) Those coordinates are then used to position individual radiosources along the instantaneous active IFT, determined by one of the three methods described in next sections, at $f=f_{ce}$ (assuming CMI emission at $f_{ce}$) as a function of time. Each t-f pair of coordinates thus corresponds to one SIII jovicentric westward longitude (hereafter longitude) of the host flux tube and one altitude along it. (3) Assuming straight line propagation from the source to the observer, we finally compute $\theta(f) = ({\bf k}, {\bf B})$ for northern sources and $\theta(f) = ({\bf k}, {\bf -B})$ for southern ones, so that both quantities can be easily compared. 

The magnetic field is modeled by using the JRM09 internal field model \cite{Connerney_GRL_18} complemented by an up-to-date model of current sheet \cite{Connerney_JGR_20} (hereafter C20). We note that the effect of the latter on the internal magnetic field at the investigated altitudes is almost negligible. The final uncertainty on $\theta$ primarily results from the uncertainty of the original fit, and from the uncertainty on the position of the active IFT.

\subsection{Straight line wave propagation}

The straight line propagation hypothesis neglects wave refraction near the source and/or along the ray path. We discuss this assumption in more details below. \citeA{Galopeau_JGR_16} investigated the shape of the Io-DAM emission cone at $f=22$~MHz as seen from the Earth. To account for refraction effects near the source, the authors chose to settle the cone axis along the opposite direction of the magnetic field gradient (-$\nabla {\bf B}$) rather than along the magnetic field vector (the tilt between both directions typically reaches a few degrees at high latitudes). With this convention, they found that the Io-DAM emission cone is significantly flattened in the direction of the magnetic field vector ($i.e.$ toward the equator) at the investigated frequency, as the result of wave refraction near the source. Nonetheless, their analysis used several important assumptions which imply significant uncertainties and/or possible biases in the calculation of the aperture angle and of the azimuth of the emission cone. The authors for instance used the low order O6 magnetic field model and assumed a fixed equatorial lead angle $\delta = 20^\circ$, which is much larger than the values predicted by the updated models presented in next section. In their analysis, the authors also defined Io-A, B, C and D source regions from large rectangular boxes in the classical diagram showing the Io-DAM occurrence probability as a function of the observer's Central Meridian Longitude (CML) and Io's phase, while such regions display in practice a more complex shape and each encompass a variety of emissions per category (multiple arcs, time-variable $\delta$) as shown by \citeA{Marques_AA_17}. Hereafter, we keep the definition of $\theta$ relative to {\bf B}, both because CMI intrinsically amplifies waves with respect to {\bf B} and to facilitate comparisons with past studies, the small angular separation (${\bf B}$,-$\nabla {\bf B}$) being left as a possible source of noise. We also neglect azimuthal variations of $\theta$, as we restrict our analysis to well-identified individual arcs corresponding to azimuthal directions where refraction near the source is expected to be low ($\pm\sim90^\circ$ from the magnetic field vector ${\bf B}$). We further discuss these hypotheses in section \ref{discussion}.


\subsection{Energy of the source electrons}

Finally, we adopted the simple formalism of \citeA{Hess_GRL_08} to infer from $\theta$ the velocity, and therefore the kinetic energy, of source electrons accelerated by the underlying Io-Jupiter Alfv\'enic current system : 

\begin{equation}
\theta = \arccos{(\frac{v}{c}\frac{1}{\sqrt{1-f_{ce}/f_{ce,max}}})}
\label{eq1}
\end{equation}

where $v$ is the characteristic velocity of resonant electrons, $c$ the speed of light and $f_{ce,max}$ is the maximum gyrofrequency at the field line atmospheric footprint. 

This equation relies on the assumptions that the refractive index at the source is $\sim1$ and that waves are amplified at $f=f_{ce}$ through loss cone-driven CMI along a prominent resonance circle in the velocity plane tangent to the loss cone. The latter hypothesis appears to be oversimplified when compared to typical loss cone electron distribution functions sampled by Juno near Io-DAM radio sources, for which a variety of possible CMI resonance circles leading to positive wave growth rate co-exist, without being necessarily tangent to the loss cone \cite{Louis_AGU_20}. We nonetheless consider equation \ref{eq1} as reasonable enough to estimate the typical velocity of electrons resonating with Io-DAM waves.

\section{Updated models of lead angle}
\label{lead_angle}

As mentioned in the introduction, the accurate positioning of the active IFT - step (2) in previous section - is central in the determination of $\theta(f)$ and appears to be the main source of uncertainty, well above that resulting from the t-f fitting of Io-DAM arcs. To accurately set the active IFT, we developed and tested three independent methods. 

The first, indirect, method dealt with in this section, is based on models of equatorial lead angles $\delta$, updating those calculated by \citeA{Hess_PRE8_17}, where $\delta$ is the longitudinal difference at the equator between the magnetic flux tube mapping to the Io's MAW spot in one hemisphere and the instantaneous position of the moon. For this purpose, we used three sets of coordinates of Io's MAW spot which were magnetically projected onto the equator with the JRM09+C20 model. The first set of coordinates corresponds to the average position of Io's MAW UV spot at 900~km altitude derived by \citeA{Bonfond_JGR_09} (hereafter B09) from a series of HST images acquired in 2007. We alternately used an updated set of coordinates fitting the Io's MAW UV footprint from the more recent study of \citeA{Bonfond_JGR_17} (hereafter B17), derived from an extended HST dataset ranging from 1997 to 2014 (therefore including the B09 data). Finally, we also used model coordinates of Io's MAW spot, whose longitude has been most recently derived by \citeA{Hinton_GRL_19} (hereafter H19) directly from a model of Alfv\'en wave propagation in the Io torus. The associated latitude was interpolated on the JRM09-derived Io footprint. 

\begin{figure*}[ht!]
\centering
\noindent\includegraphics[width=30pc]{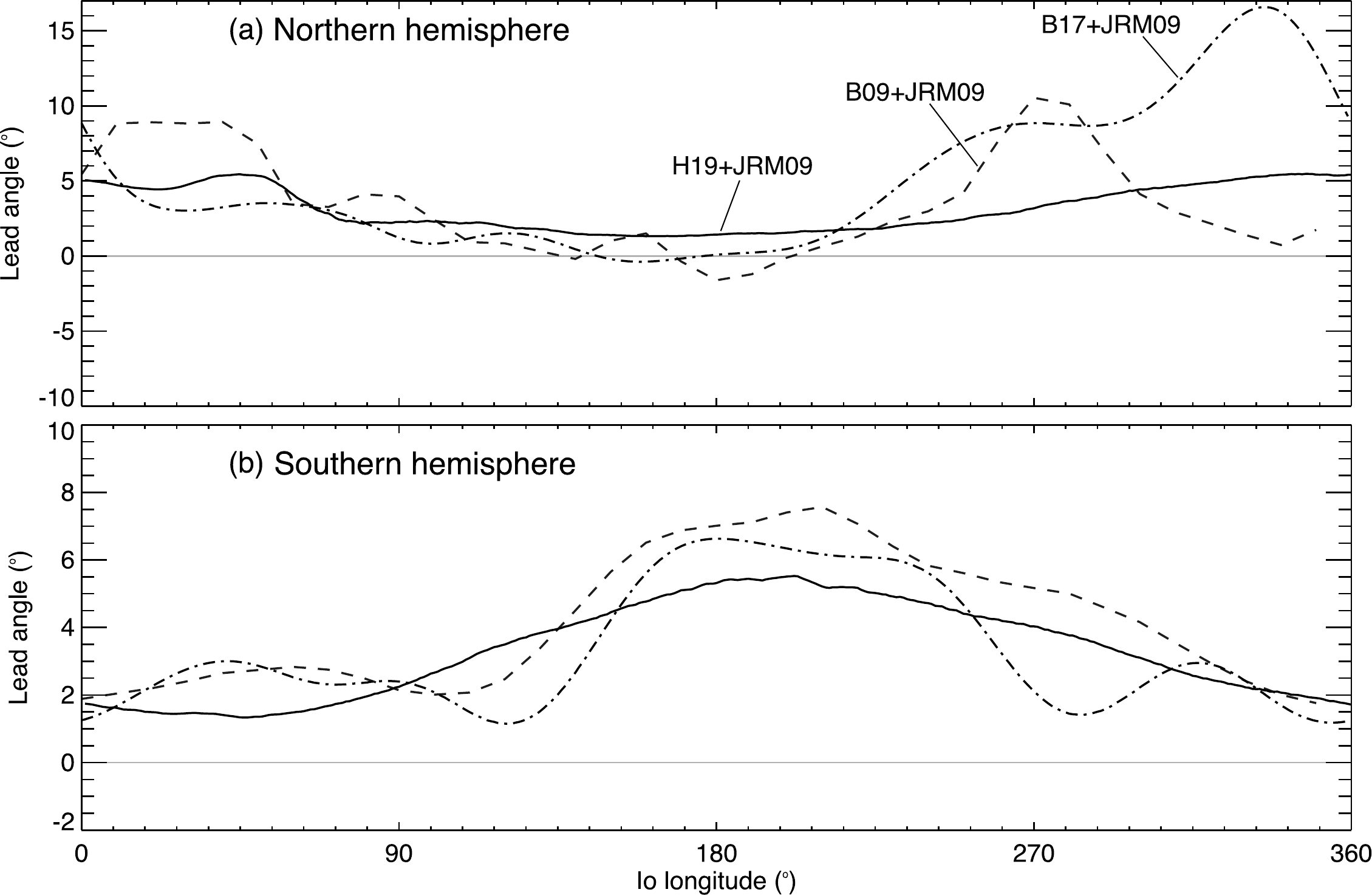}
\caption{Models of equatorial lead angles $\delta$ in the northern (top) and in the southern (bottom) hemispheres defining the location of the active IFT with respect to the instantaneous IFT, as derived from the three sets of coordinates of Io's MAW spot provided by \cite{Bonfond_JGR_09,Bonfond_JGR_17,Hinton_GRL_19}.}
\label{fig1}
\end{figure*}

The obtained three models of $\delta$ are displayed in Figure \ref{fig1} as a function of Io's longitude. They display similar smooth trends in both hemispheres. In the north, they vary from $\delta\sim0$ near $180^\circ$ longitude to $\delta\sim10^\circ$ in the $270-50^\circ$ range (except for the B17 model for which $\delta$ reaches $16^\circ$ near $330^\circ$ longitude, where Io's UV footprint is more difficult to track in HST images). In the south, they vary from $1^\circ$ to $7.5^\circ$, with a well-defined maximum near $\sim200^\circ$. Supplementary Figure S1 compares those lead angle models to those previously derived by \citeA{Hess_PRE8_17}. Overall, the new models display much less variability and predict positive (physical) values of $\delta$ (except for B09 $\delta$ in the north, where $\delta$ briefly reaches a $-1.5^\circ$ local minimum at $180^\circ$ longitude). Comparing the three models, the values of $\delta$ differ by only a few degrees, generally less than $\pm2^\circ$, up to localized $5-10^\circ$ differences in the north. 

In the following, the three models will generally be used together for the determination of the active IFT, their difference providing an estimate of the typical uncertainty on $\delta$. 

\section{Simultaneous radio/UV observations}
\label{radio_uv}

Radio and UV auroral emission processes being tied to the same accelerated electron population, the radio emission corresponding to an isolated Io-DAM arc (or to the main arc whenever a series is present) was long thought to be colocated with the Io MAW UV spot. Multiple Io-DAM arcs observed intermittently were also suspected to correspond to Io secondary (RAW) UV spots, in agreement with the expected topology of the Alfv\'enic current system \cite{Gurnett_JGR_81,Hess_JGR_10}. \citeA{Hess_PSS_10}, for instance, fitted multiple Io-DAM arcs and found differences of lead angle consistent with the typical average interspot longitude interval of $6^\circ$ within Io's UV footprint emission. However, to our knowledge, the direct correspondence between radio and UV multiple emissions has not been unambiguously established to date.

\subsection{The case of day 2017-01-27}

\begin{figure*}[ht!]
\centering
\noindent\includegraphics[width=30pc]{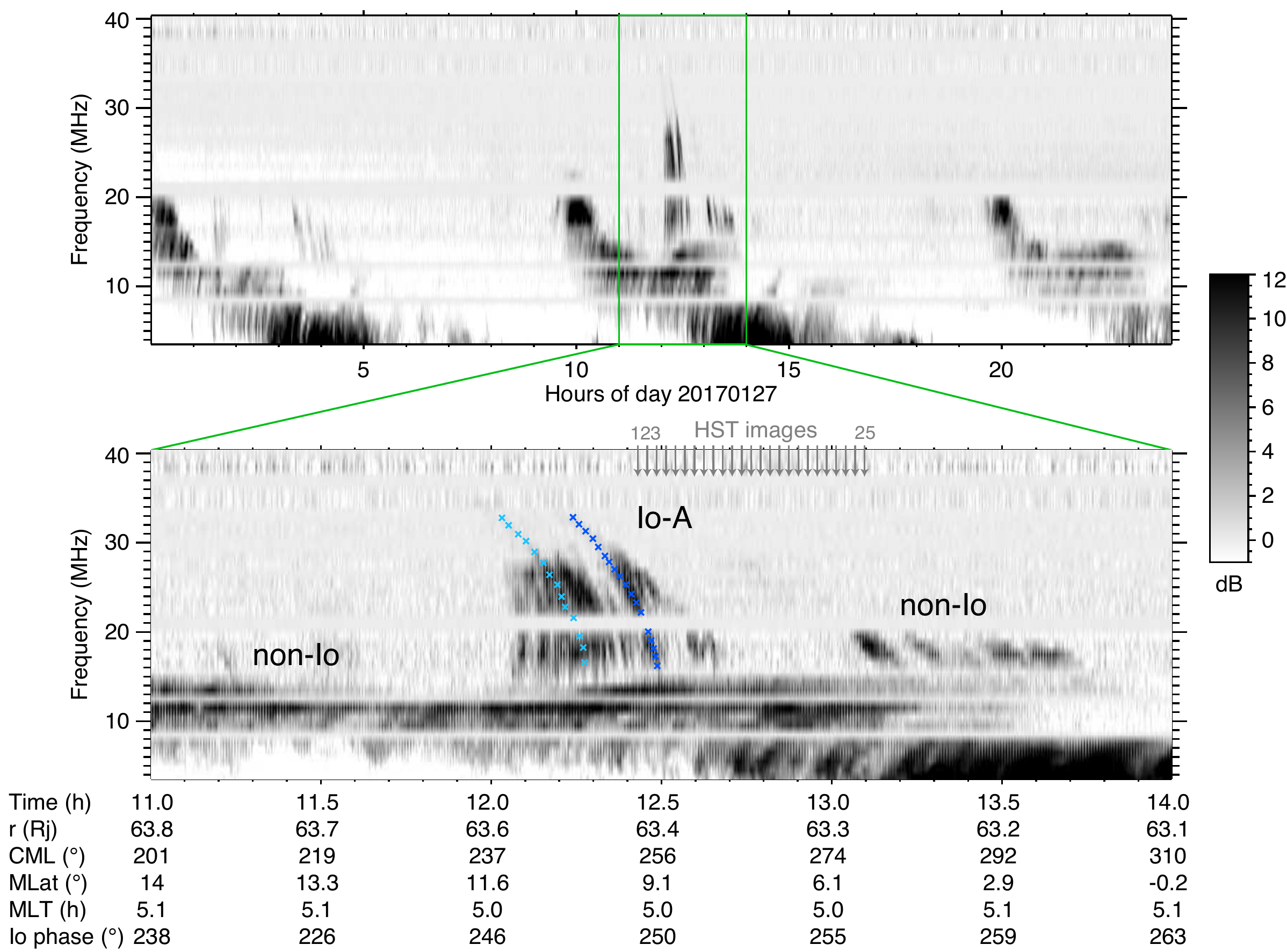}
\caption{(Top) Juno/Waves observations of Jupiter between 3 to 41~MHz on 27th Jan. 2017. A bright double arc structure, reaching frequencies as high as $\sim$34~MHz, corresponds to Io-A emission. (Bottom) Focus on the Io-A DAM interval, which displays a complex series of arcs, on top of non-Io emissions. The two most intense arcs have been fitted by blue crosses (the faint, high frequency, portion of the fitted emissions is more visible on the top panel). The observing times of $25\times100$~sec long HST images are indicated by gray arrows.}
\label{fig2}
\end{figure*}

In this framework, we searched for simultaneous radio/UV observations of auroral emissions induced by Io, allowing for real-time determination of the active IFT, by cross-matching catalogs of Io-DAM events recorded by Juno/Waves \cite{Louis_JGR_21,2021_Juno_Waves_catalog} and the NDA \cite{Marques_AA_17} on the one hand, and of Io UV footprints detected in HST images on the other hand throughout a 4 years-long interval ranging from mid-2016 to mid-2020. We found a single event (only), whose radio measurements are displayed in Figure \ref{fig2}. 

On 27th Jan. 2017, between 12$:$00 and 12$:$40~UT (hereafter, all times have been light-time corrected and correspond to times measured at Jupiter), Juno/Waves observed two, strikingly similar, bright radio arcs reaching frequencies as high as $\sim34$~MHz (Figure \ref{fig2}, top), among a series of fainter ones confined below $\sim$30~MHz (Figure \ref{fig2}, bottom). Juno was located at magnetic latitudes near $+10^\circ$ and the arcs displayed a characteristic vertex-late curvature which enabled us to identify them as Io-A emissions from the northern magnetic hemisphere. This identification was confirmed by checking classical CML-Io phase diagrams mapping the occurrence of Io-DAM emissions, such as Figure 7 of \cite{Marques_AA_17}, with the online Jupiter probability tool available at \url{https://jupiter-probability-tool.obspm.fr}. 

We then fitted the main (latest observed) arc and the secondary (earliest) one by continuously tracking local intensity maxima as a function of frequency, as indicated by dark and blue crosses, respectively. Supplementary Figure S2 displays flux density time series at different frequencies, ranging from 16 to 33~MHz, which better show these intensity maxima and illustrate the relevance of the fit.

\begin{figure*}[ht!]
\centering
\noindent\includegraphics[width=30pc]{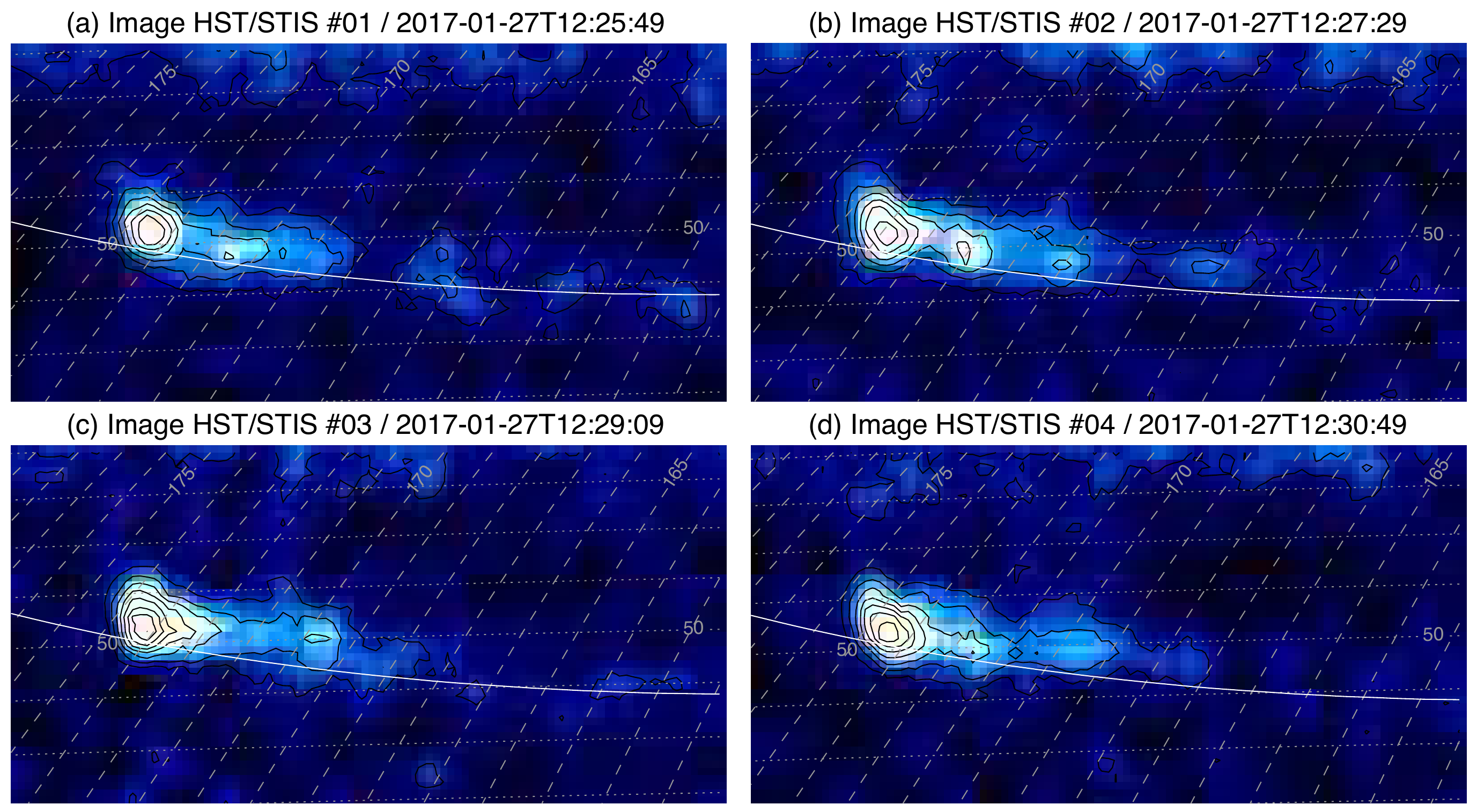}
\caption{HST/STIS FUV images of Jupiter's northern aurorae numbered 1-4 (according to the labelling of Figure \ref{fig2}), each corresponding to a 100-sec long exposure, acquired on day 2017-01-27. Intensity iso-contours of N ($\ge$1) standard deviations above the background level are shown with solid black lines, on top of planetocentric latitudes (dotted gray lines), longitudes (dashed gray lines) and JRM09-derived footpath of Io (solid white line) at 900~km altitude.}
\label{fig3}
\end{figure*}

On the same day, between 12$:$25 and 13$:$06 UT, HST observed the northern auroral region with a FUV STIS long exposure image, which we sequenced into $25\times100$~sec long sub-exposures labelled 1 to 25 at times indicated as grey arrows in Figure \ref{fig2}. Images 1-4 were acquired strictly simultaneously to the latest portion of the Io-A main arc. Figure \ref{fig3} displays HST images 1-4, together with a grid of planetocentric coordinates (gray lines) and the JRM09-derived Io footpath (solid white line) at 900~km altitude. Black solid lines mark intensity iso-contours every standard deviation ($\sigma$) above the background level. The complete, animated, set of images is provided as supplementary animation S3. Overall, the northern Io footprint emission displayed a persistent, bright, main spot (on the left-hand side), straightforwardly identified as the MAW spot. Io was indeed near $\sim180^\circ$ longitude during the STIS observation, namely near the northern edge of the torus, a position for which no precursor/leading spot is expected. The UV images also revealed a series of secondary spots lagging the main one at lower longitudes, the first of which was especially visible on images 1-2 and 4. 

\subsection{Association of both radio arcs with the MAW and TEB UV spots}

\begin{figure*}[ht!]
\centering
\noindent\includegraphics[width=15pc]{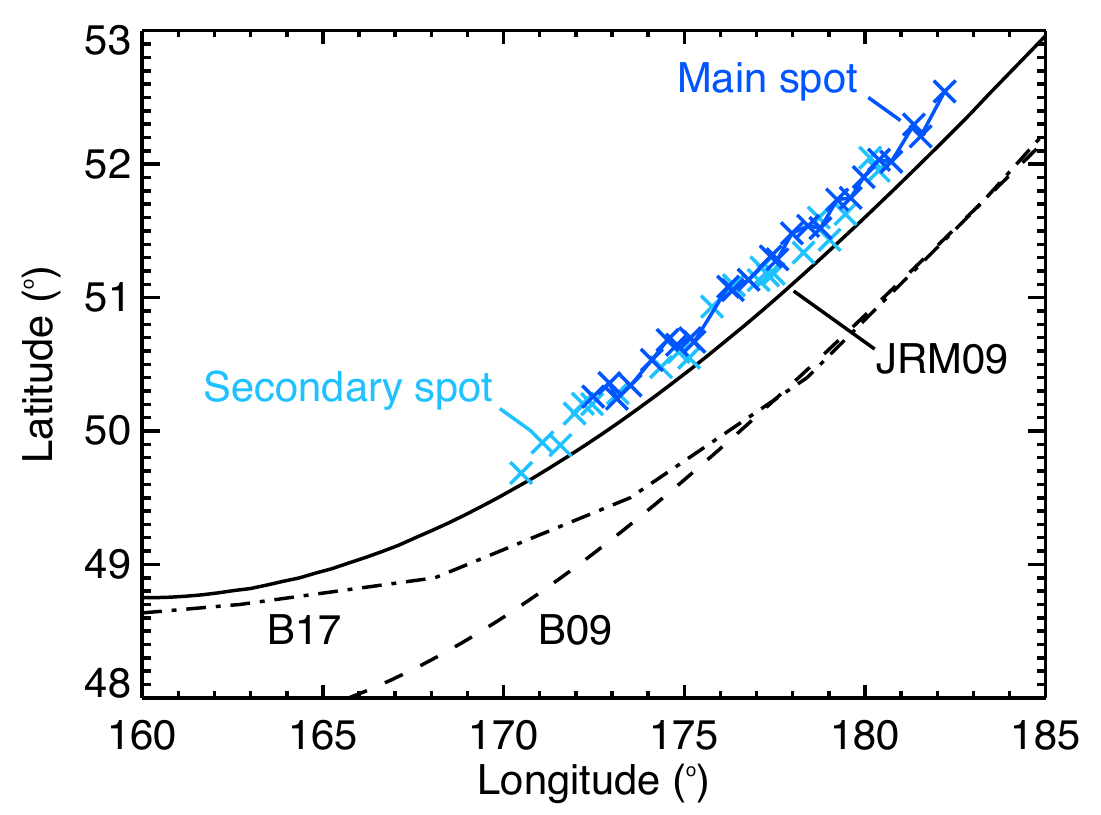}
\caption{The dark (light) blue crosses indicate the planetocentric latitude and longitude (at 900~km above the 1-bar level) of the main and secondary UV spots which could be automatically tracked from the 25 HST/STIS images, respectively. The solid, dashed and dotted lines map the footpath of Io predicted by JRM09 (H19), B09 and B17, respectively. Whatever the model chosen, $\delta$ remains $\le1.5^\circ$ for a given longitude.}
\label{fig4}
\end{figure*}

We automatically tracked the coordinates of the MAW and first secondary UV spots on all the 25 images at an altitude of 900~km, as displayed in Figure \ref{fig4}. Both spots followed a very similar trajectory, very close to the JRM09-derived Io footpath (solid line), and farther from the Io's MAW trajectory from B09 (dashed) and B17 (dotted-dashed). We then quantitatively compared the coordinates of the MAW UV spot to the H19, B09 and B17 expected ones at each time, for a given longitude of Io. The observed spot was located in average (median value) at $+0.32^\circ$ latitude from the H19 prediction and farther from the B17 and B09 one, while lagging in longitude the H19, B09 and B17 predictions by $-1.1^\circ$, $-2.3^\circ$ and $-1.7^\circ$ respectively. We attribute the $+0.32^\circ$ latitudinal difference to the uncertainty of our polar projection at 900~km. For comparison purposes, we checked that a polar projection at an altitude of 300~km instead yields a MAW footprint shifted by $+0.74^\circ$ in latitude. This suggests that the effective altitude of the observed Io auroral footprint was slightly larger than 900~km for this event. We then consider the longitudinal shift, compared to the $\sim$$1.4^\circ$ longitudinal width at half maximum of the spot, as highly significant. Such a lag turns to be the main source of uncertainty on the determination of $\theta(f)$, as we will see later on, and illustrates the value-added of simultaneous UV observations to correctly position the instantaneous active IFT.

The UV observations bring additional important informations. The median difference in longitude between the MAW and the secondary spot is $1.71^\circ$. This value is closer to the longitudinal difference between the MAW northern spot and the Transhemispheric Electron Beam (TEB) spot associated with the MAW southern spot, as predicted by H19, than with the first RAW northern spot, expected to lag the MAW spot by $\sim$5$^\circ$ (a spot is sometimes visible at such a longitudinal distance in Figure \ref{fig3} and in supplementary animation S3 but was not considered further in our analysis). Conversely, the $\sim0.2~h$ time delay between the low frequency edges of the two fitted Io-DAM arcs corresponds to a longitudinal difference of $\sim2^\circ$ along the northern Io footpath. This fair agreement supports our assumption that the main (secondary, respectively) Io-DAM arc is driven by radiosources hosted by a flux tube mapping to the MAW (TEB, respectively) UV spot and brings the first evidence that up-going TEB in one hemisphere can drive radio emission in the other hemisphere. Furthermore, we notice that the longitudinal extent of the main 'active' UV region ($\ge1\sigma$ contour) reaches 6 to 9$^\circ$ in the 4 images of Figure \ref{fig3}. This value again fairly matches the $\sim$0.7~h temporal duration of the whole series of Io-A arcs, which transposes into a $\sim$9$^\circ$ longitudinal extent along the northern Io footpath. This correspondence suggests that multiple radio arcs are likely associated with multiple UV sub-structures between the MAW and RAW spots. Supplementary Figure S2 shows at least 9 distinct successive Io-A arcs more or less regularly spaced. The fact that the UV tail appears to be sometimes (much) more elongated (see supplementary animation S3), while the series of radio arcs is well clustered may result from a better sensitivity of HST/STIS to observe $H_2$ band emission than of Juno/Waves to detect Io-DAM radio waves. Conversely, the high temporal cadence of radio observations appears as a powerful mean to track a variety of small-sized active flux tubes (Figure S2) undetectable at the spatial resolution of UV spectro-imagers such as STIS. 

In summary, this simultaneous radio/UV observation provided evidence that the main and secondary radio arcs map to the main (MAW) and secondary (TEB) UV spots, respectively.

\subsection{Characteristics of the Io-A main arc}

\begin{figure*}[ht!]
\centering
\noindent\includegraphics[width=30pc]{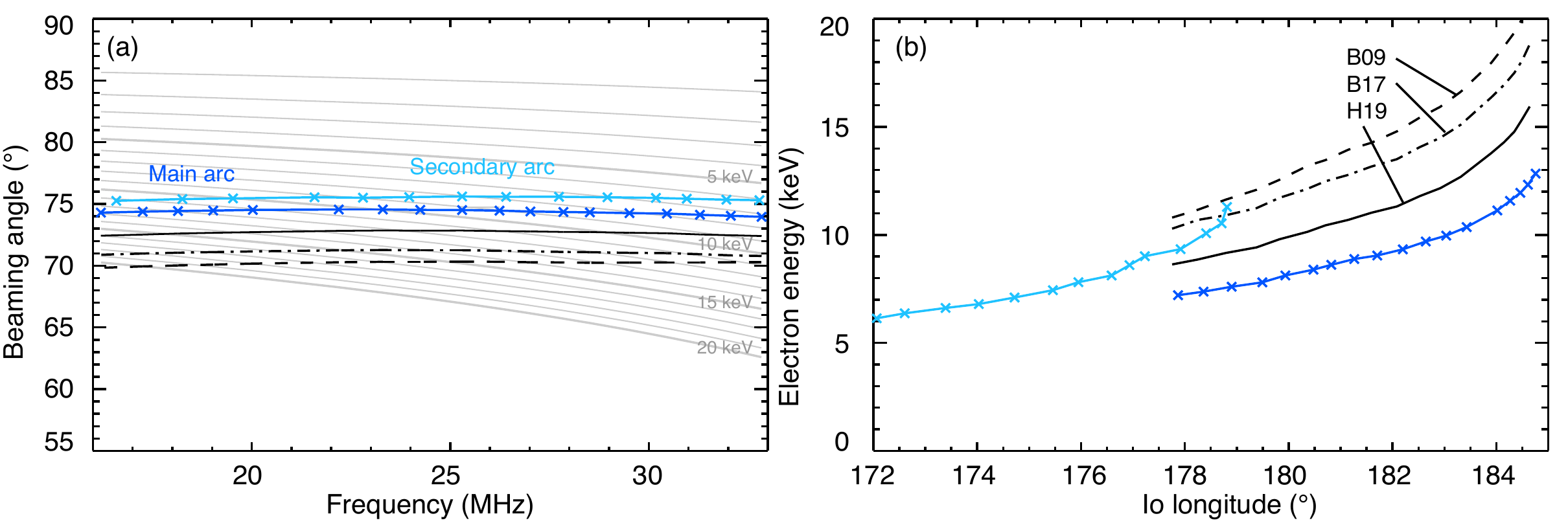}
\caption{(a) Radio emission angle at the source $\theta$ as a function of increasing frequency (decreasing altitude along the flux tube). Dark (light) blue crosses correspond to emission angles computed for the Io-DAM main (secondary) arc, as fitted in Figure \ref{fig2}, bottom, once associated with the MAW (TEB) UV spot, as fitted in Figure \ref{fig4}. Both Io-A arcs display very similar emission angles, although slightly larger for the secondary one. The solid, dashed and dotted black lines provide values of $\theta(f)$ when positioning the active IFT with model lead angles H19, B09 and B17 (see section \ref{lead_angle}). The series of gray lines plot the expected theoretical value of $\theta(f)$ computed from equation \ref{eq1} for a time-variable flux tube associated with the MAW UV spot for electron kinetic energies ranging from 1 (top curve) to 20~keV (bottom curve). (b) $E_{e-}$ of CMI-unstable electrons derived from equation \ref{eq1} inferred from the values of $\theta(f)$ displayed on panel (a). Both Io-A arcs again display similar trends although with slightly lower energies for the secondary arc.}
\label{fig5}
\end{figure*}

Once having identified the two flux tubes hosting the radiosources responsible for the main and secondary Io-A arcs - step (2) in section \ref{beaming} - it is straightforward to derive $\theta$ for the t-f coordinates fitting each radio arc as a function of frequency, as displayed in Figure \ref{fig5}, left. The main radio arc yields roughly constant emission angles of $\theta(f)\sim75^\circ$ (dark blue crosses). The temporal uncertainty on the fit is estimated by the $\sim$0.03~h half full width at half maximum of the arc which yields a typical error on $\theta$ of $\sim$$0.3^\circ$. It should be noted that this error is overestimated, as the temporal thickness of the arc results from the convolution of the width of the emission cone with the spatial extent of the active region (see e.g. \cite{Lamy_JGR_08b}), both of which are neglected here.

Alternately, moving the field line footprint by $1^\circ$ longitude along the tracked UV footpath (dark blue crosses in Figure \ref{fig5}) yields a typical error on $\theta$ of $\sim$$1.5^\circ$. For comparison purposes, the set of black lines indicate emission angles derived from the three lead angle models described at section \ref{lead_angle}, whose predictions were shown above to significantly lag the MAW UV footprint. They yield lower values of $\theta(f)$, down to $-4^\circ$ for the B09 model with respect to the dark blue crosses.

The set of gray lines superimpose theoretical values of $\theta(f)$ derived from equation \ref{eq1} with the UV-derived set of flux tubes and $f_{ce,max}$ derived at a 900~km altitude mirror point, for electron kinetic energies $E_{e-}$ ranging from 1~keV (top) to 20~keV (bottom). The observed $\theta(f)$ intercept theoretical curves corresponding to a limited range of $E_{e-}$ : from 7~keV at 33~MHz (earliest arc edge) to 13~keV at 16~MHz (latest arc edge). The observed (constant) and theoretical (decreasing) trends also clearly differ, suggesting that $E_{e-}$ was increasing with decreasing frequency (increasing altitude) and/or with increasing time. This trend can be better seen in Figure \ref{fig5}, right, which directly displays $E_{e-}$ computed from $\theta(f)$ as a function of Io's longitude (increasing time). $E_{e-}$ smoothly increases from 7~keV at $178^\circ$ to 13~keV at $185^\circ$. The set of black lines indicate for comparison $E_{e-}$ derived from the three lead angle models discussed in section \ref{lead_angle}. All of them show the same trend as a function of Io's longitude, although corresponding to larger $E_{e-}$, up to $+6$~keV for the B09-derived values.

This case study likely yields the most accurate determination of $\theta(f)$ for an Io-DAM arc to date. While the obtained values are consistent with those published previously, they also significantly differ from the expected decrease as a function of magnetic field amplitude, or equivalently as a function of frequency, from equation \ref{eq1} for loss cone-driven CMI emission assuming a constant electron velocity. Instead, the kinetic energy of electrons driving the radiation appears to vary with time (Io's longitude) and/or frequency (altitude) along the active IFT, here up to a factor of 2 over the 14~min duration and 17~MHz bandwidth of the Io-A main arc.

\subsection{Characteristics of the Io-A secondary arc}

Looking now at the characteristics of the secondary radio arc associated with the TEB UV spot, light blue crosses indicate emission angles strikingly similar to those of the main radio arc, although larger by $\sim1^\circ$. This in turn yields electron energies increasing from 6 to 12~keV from Io's longitude evolving from 172 to 177$^\circ$. These values strikingly compare to those inferred above for the main Io-A arc. Moreover, as the TEB northern UV spot displays brightnesses half those of the MAW northern UV spot, comparable electron energies imply that the electron flux responsible for the TEB UV spot is likely half that feeding in the MAW UV spot. 


\section{Multi-point radio observations}
\label{radio_radio}

As simultaneous radio/UV observations are quite rare, we alternately attempted to benefit from bi-point simultaneous radio observations of Io-DAM arcs to constrain the position of radiosources by comparing their emission angles measured in different directions by two observers. To search for such events, we cross-matched catalogs of Io-DAM events recorded by Juno/Waves and the NDA since mid-2016. We identified two episodes during which Io-DAM arcs were observed partly simultaneously by both instruments. 


\subsection{The case of day 2018-03-05}

\begin{figure*}[ht!]
\centering
\noindent\includegraphics[width=30pc]{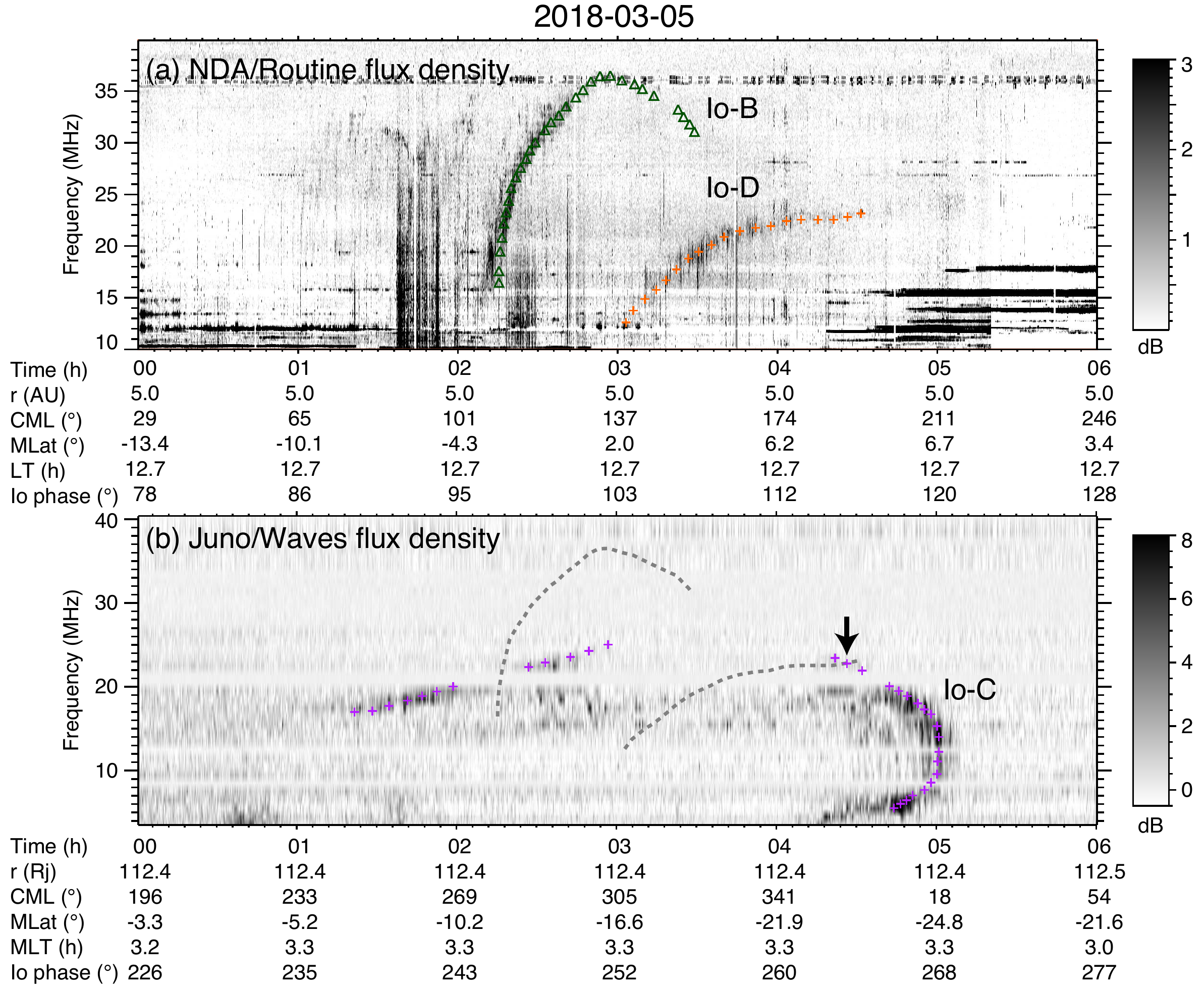}
\caption{Radio observations of Jupiter acquired simultaneously with (a) the NDA (once corrected for light-time travel) and (b) Juno/Waves on 5th Mar. 2018. Both panels display dynamic spectra of flux density which include (a) Io-B and D and (b) Io-C emissions, fitted with orange triangles and red crosses, respectively. The Io-B and D arcs fitted on panel (a) have been replicated on panel (b) with gray dashed lines. There, the Io-D dashed line intercepts the Io-C arc, fitted with purple crosses, near 04$:$30 and $\sim$20~MHz, as marked by the black arrow.}
\label{fig6}
\end{figure*}

Figure \ref{fig6} shows radio observations of Jupiter acquired by the NDA (once corrected for light-time travel) and Juno/Waves on 5th Mar. 2018. For simplicity, Figure \ref{fig6}a displays the total flux density retrieved from NDA/Routine LH and RH polarized measurements. From 02$:$00 to 04$:$30~UT, from a near-Jovian noon near-equatorial observing direction, the NDA observed two characteristic Io-DAM arcs, respectively RH and LH polarized, originating from the northern and southern hemispheres, and therefore identified as Io-B and Io-D emissions. Their vertex-early curvature indicates that they were observed from the east. The colored symbols on Figure \ref{fig6}a materialize the fit of both arcs. Meanwhile, between 01$:$00 and 05$:$00~UT, Juno/Waves observed an Io-C emission from southern magnetic latitudes in the early morning sector. The vertex-late shape of the outer main arc, fitted by purple crosses on Figure \ref{fig6}b, indicates that the southern active IFT was, here, observed from the west. The weaker, inner, secondary Io-C arc is left aside from the analysis below. The NDA-fitted Io-B and D arcs have been replicated on the Juno/Waves dynamic spectrum with gray dashed lines. There, the Io-D curve intercepts the Io-C arc near 04$:$30~UT and $\sim$20~MHz (black arrow). At this time (IFT position) and frequency (source altitude), the same emission cone was therefore simultaneously observed in two different directions, roughly symmetrically eastward and westward from the local magnetic field vector. 

\begin{figure*}[ht!]
\centering
\noindent\includegraphics[width=30pc]{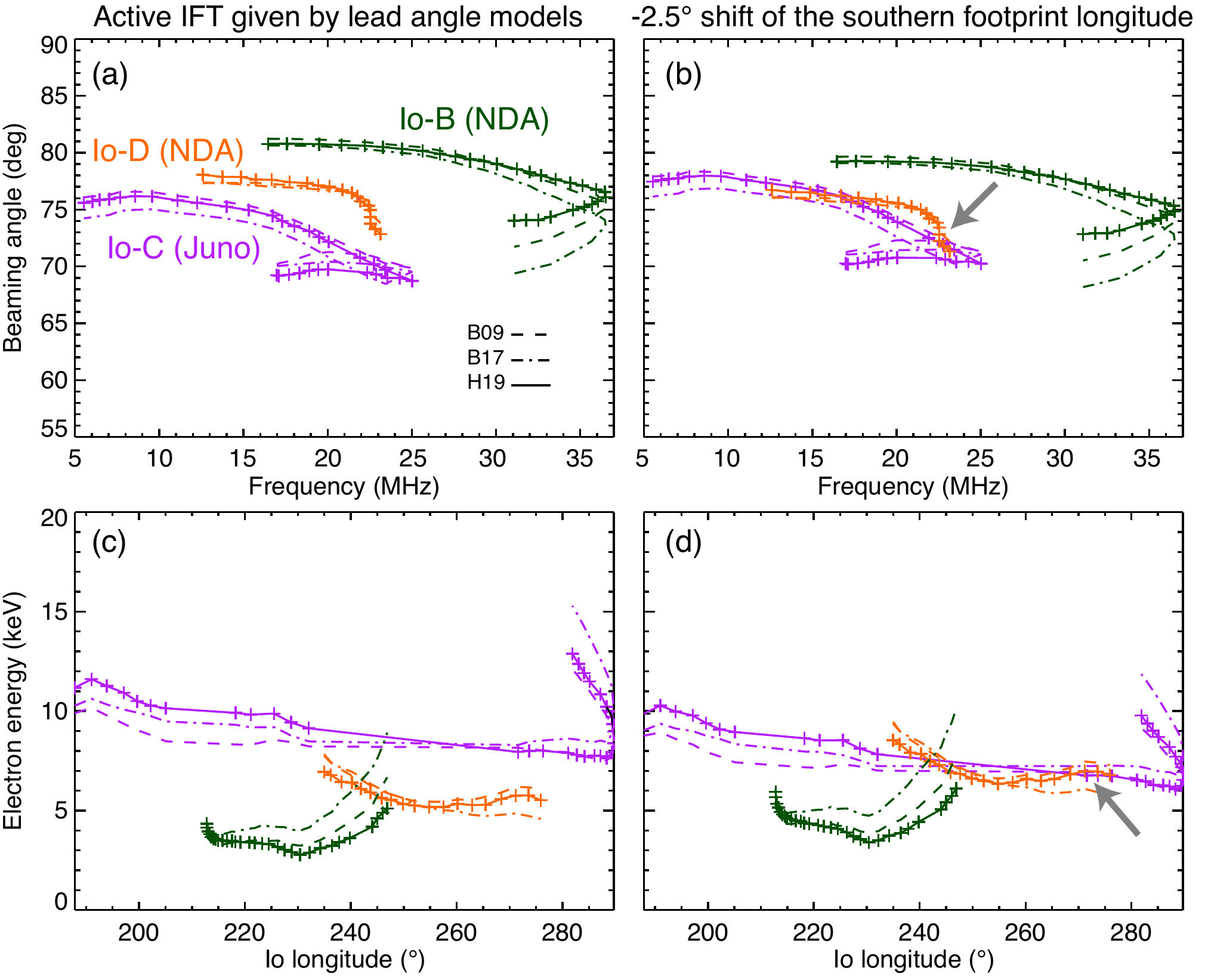}
\caption{(a) Radio emission angle at the source $\theta$, plotted as a function of increasing frequency (decreasing altitude) for the Io-D and Io-C emissions fitted on Figure \ref{fig6}, here displayed by red and purple crosses, respectively. Each set of three lines displays the values of $\theta$ obtained from the H19, B17 and B09 lead angle models. (b) Same as (a) but correcting the modeled longitude of the southern IFT footprint by a $-2.5^\circ$ shift. (c) CMI-unstable electron energies inferred from panel (a) through equation \ref{eq1}. (d) Same as (c) but for the values of $\theta$ derived in panel (b). On panels (a) and (c), the two sets of red and purple curves do not intersect. On panels (b) and (d), they fairly match and coincide at $\sim$20~MHz (gray arrows), where the same radio source was simultaneously observed by the NDA and Juno/Waves.}
\label{fig7}
\end{figure*}

Using the H19, B17 and B09 equatorial lead angle models to position the southern active IFT as a function of time, we determined $\theta(f)$ and $E_{e-}$ along the flux tube hosting both the Io-D and C southern emissions. These are plotted by sets of red and purple lines in Figure \ref{fig7}a and c, respectively as a function of frequency and Io's longitude. The two sets of values do not overlap with each other, contrary to expectations for an azimuthally symmetric emission cone. We have thus tested a range of longitudes of the southern IFT slightly shifted from the modeled ones. In Figure \ref{fig7}b and d, the values of $\theta(f)$ and $E_{e-}$ correspond to a $-2.5^\circ$ longitudinal correction of the southern footprint, which we showed to be a realistic value (for the northern hemisphere) in section \ref{radio_uv}. The red and purple sets of curves now fairly match with each other, achieving symmetrical emission angles and similar electron energies at the intersection of Io-D and C arcs (gray arrows). The northern IFT, hosting the Io-DAM sources responsible for the Io-B arc fitted in Figure \ref{fig6}a, was similarly corrected by propagating the $-2.5^\circ$ shift of the southern footprint longitude, yielding a $-1.4^\circ$ shift of the northern footprint longitude. The Io-B parameters obtained with (without, respectively) this correction are displayed in green in Figure \ref{fig7}b and d (a and c, respectively).

The overall agreement between the Io-D and C-derived $\theta(f)$ and $E_{e-}$ again supports a variation as a function of frequency (altitude) and/or time (Io's longitude). A means by which one can disentangle the two variations is provided by the shape of Io-DAM arcs. The curvature of the Io-C arc was such that, between 04$:$40 and 05$:$00~UT, different frequencies were observed simultaneously. Over this time interval, two active radio sources were thus detected simultaneously at two different altitudes along the same flux tube. Figure \ref{fig7}d indicates that the electron energies were larger at lower frequencies (higher altitudes) by a few keV, proving the variation of $E_{e-}$ as a function of altitude whereas it was generally assumed to be constant along the active IFT in past studies. The same arc was also curved enough so that frequencies between 17~MHz and 23~MHz were sampled at two different times, probing the same altitude for two different positions of the active IFT. This time, $E_{e-}$ was larger by a few keV during the first half of the arc. This decrease of $E_{e-}$ with time is supported by the overall agreement between the purple (Io-C) and orange (Io-D) curves, despite the arcs probed different frequencies (altitude) out of their intersection point.

Altogether, the analysis of the Io-C and D arcs supports a variation of $E_{e-}$ as a function of time, or equivalently as a function of the position of Io within the torus, as already suggested by \citeA{Hess_PSS_10}.

\subsection{The case of day 2017-01-29}

\begin{figure*}[ht!]
\centering
\noindent\includegraphics[width=30pc]{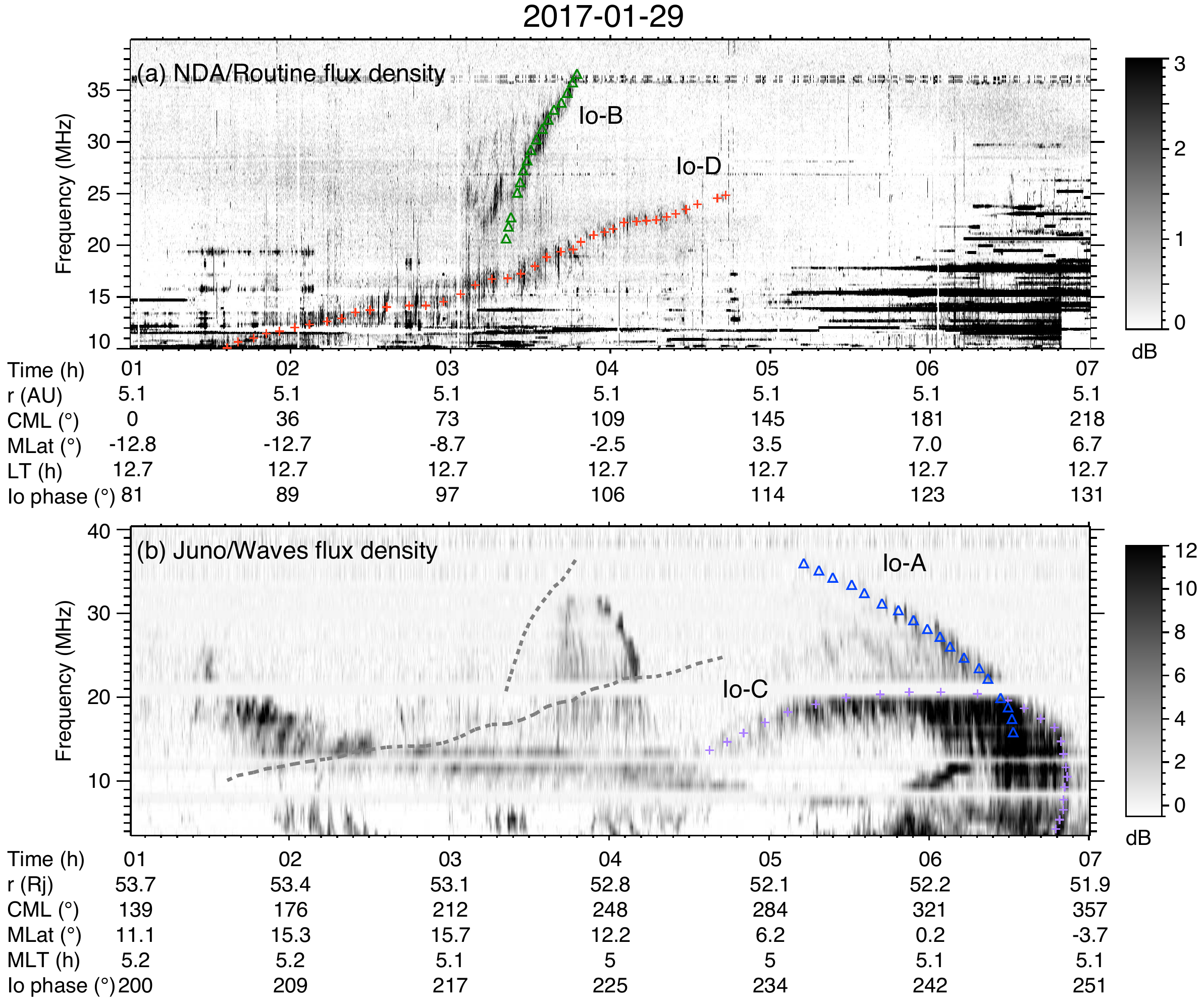}
\caption{Radio observations of Jupiter acquired simultaneously with (a) the NDA (once corrected for light-time travel) and (b) Juno/Waves on 29th Jan. 2017. Both panels display dynamic spectra of flux density which include (a) Io-B and D and (b) Io-A and C emissions, fitted with colored triangles and crosses for northern and southern emissions, respectively. The Io-B and D arcs fitted on panel (a) have been replicated on panel (b) with gray dashed lines. While the latest portion of the Io-D arc fit was observed simultaneously to the earliest portion of the Io-C arc, both were separated by a few MHz and did not intersect each other.}
\label{fig8}
\end{figure*}

Figure \ref{fig8} shows another interesting example of Io-DAM emissions simultaneously observed by the NDA and Juno/Waves on 29th Jan. 2017. As Juno was located close to the magnetic equator, it could detect emissions from both hemispheres, such as the NDA. Specifically, Io-B (RH polarized, northern hemisphere) and Io-D (LH polarized, southern hemisphere) vertex-late arcs were observed by the NDA from 01$:$00 to 05$:$00~UT, while Io-A (northern hemisphere) and Io-C (southern hemisphere) vertex-early emissions were tracked by Juno/Waves between 04$:$30 and 07$:$00~UT. That is, the northern and southern parts of the active IFT were observed from both eastward and westward directions during a continuous $\sim$7~h-long interval ($\sim$3/4 of a Jovian rotation). The northern Io-A and B arcs were observed strictly successively, $\ge1.3$~h apart. As for the southern emissions, while the latest portion of the Io-D fit replicated on Figure \ref{fig8}b overlapped the early portion of the Io-C arc during a few minutes, they did not intersect. 

\begin{figure*}[ht!]
\centering
\noindent\includegraphics[width=30pc]{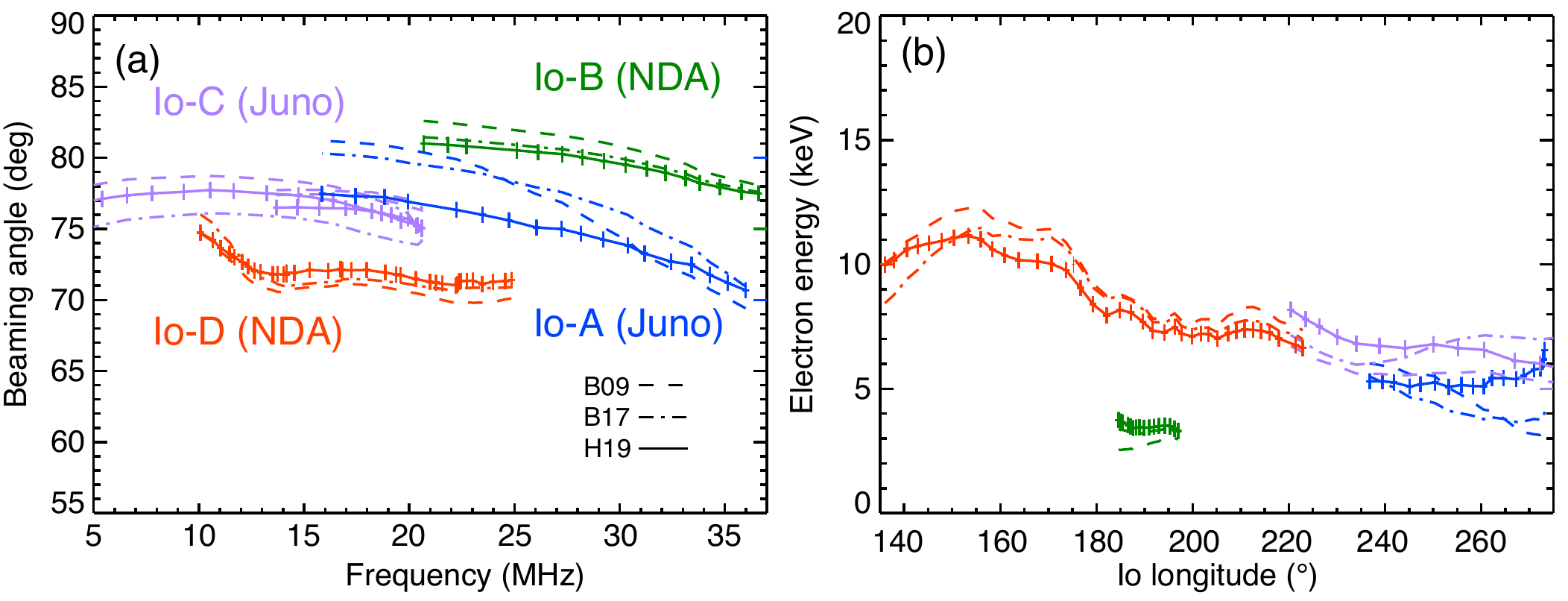}
\caption{Same as Figure \ref{fig7} for the four Io-DAM arcs recorded by the NDA and Juno/Waves on 29th Jan. 2017. (a) Emission angle and (b) inferred electron kinetic energy obtained by deriving the active IFT with lead angle models (without further correction).}
\label{fig9}
\end{figure*}

Figure \ref{fig9}a displays $\theta(f)$ for the four arcs, determined as in Figure \ref{fig7}a from the three lead angle models. The emission angles display a similar behaviour with a decrease toward high frequencies and values varying within $82-70^\circ$. Figure \ref{fig9}b plots the inferred values of $E_{e-}$, again as a function of Io's longitude. While the sources probed during the overlap of Io-D and C arcs were located at slightly different altitudes (the Io-D sources being located a few MHz above the Io-C ones, hence closer to the planet), the corresponding portions of the red and purple sets of curves connect with each other. This suggests that $E_{e-}$ was fairly estimated (and did not vary much between those altitudes), and in turn that the active IFT was correctly positioned by the lead angle models. The agreement between both sets of red and purple curves is even more striking over the full interval, drawing a smooth decreasing trend from $10-12$ to $5-7$~keV while the longitude of Io increased from $135^\circ$ to $275^\circ$. Inspecting frequencies of the Io-D and C emissions observed at two different times clearly show that $E_{e-}$ decreased by a few keV between the two events, and further illustrates that $E_{e-}$ varies with time independently of altitude. Conversely, the Io-B and Io-A arcs observed successively yield $E_{e-}$ increasing from $\sim$3 to $6-9$~keV between $185^\circ$ and $275^\circ$ of Io's longitude. While Io was travelling in the northern part of the torus during this interval, reaching its northern edge near $196^\circ$, the southern energies remained larger than the northern ones.

The results presented in this section thus confirm both the range and the t-f variability of $\theta$ and $E_{e-}$.

\section{Single-point radio observations : toward a statistical study}
\label{stand_alone}

The above method, based on updated lead angle models, can be applied to any stand-alone (either space- or ground-based) radio observation of Io-DAM arcs. As the NDA quasi-daily observes Jupiter since January 1978, it provides an ideal dataset to perform a dedicated statistical analysis. Such a study is nonetheless beyond the scope of this paper and, in this section, we simply tested this method on a NenuFAR observation of Jupiter at high sensitivity and on one Juno/Waves observation analyzed in a separate study.

\subsection{First NenuFAR observations of Jupiter}

Among the observations of Jupiter collected in support of Juno since 2019, during its early science phase, NenuFAR sampled a couple of Io-DAM arcs. The value-added of such observations resides in the fact that the large number of antennas provides a very high sensitivity, useful to track faint extensions of Io-DAM arcs undetectable by the NDA and Juno/Waves. In addition, the degree of circular polarization can be easily derived at high t-f resolution from NenuFAR/UnDySPuTeD data and turns out to be a very efficient mean to track highly circularly polarized signal (even with low flux density) embedded within the usual RFI band below 20~MHz.

\begin{figure*}[ht!]
\centering
\noindent\includegraphics[width=30pc]{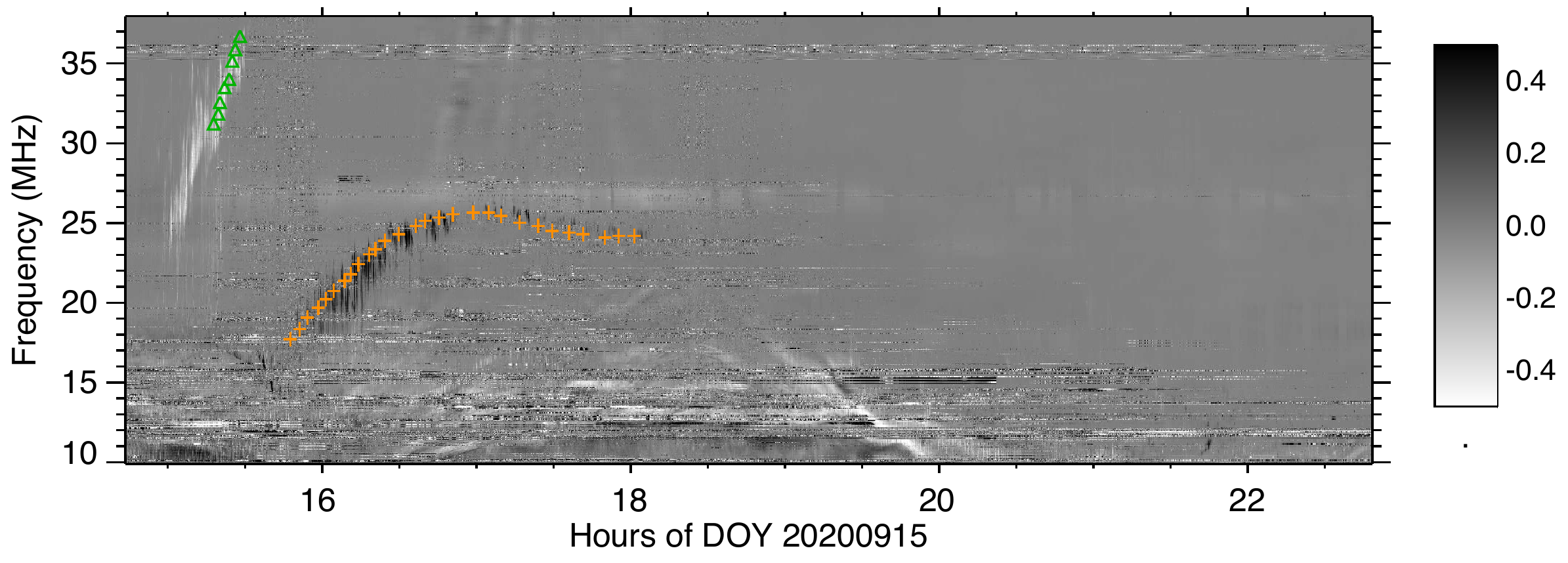}
\caption{NenuFAR dynamic spectrum of the degree of circular polarization for a Jupiter observation conducted on 15th Sept. 2020, once corrected for light-time travel. The Io-B and D arcs, respectively RH and LH polarized, have been fitted with green triangles and orange crosses.}
\label{fig10}
\end{figure*}

Figure \ref{fig10} shows an example of Io-B (RH polarized) and Io-D (LH polarized) arcs observed with NenuFAR on the late afternoon of 15th Sept. 2020, and fitted with green triangles and orange crosses. The corresponding values of $\theta(f)$ and $E_{e-}$, derived from the H19 model only (for the sake of clarity), are displayed on the summary Figure \ref{fig11} and in table \ref{tab}, where they appear to be in excellent agreement with those obtained for all the other Io-B and D arcs analyzed in this study. 

\subsection{Application to case studies investigated in other recent articles}

As mentioned in the introduction, the Io-DAM beaming was recently investigated in two other independent, parallel, studies using the JRM09 magnetic field model. 

\begin{figure*}[ht!]
\centering
\noindent\includegraphics[width=30pc]{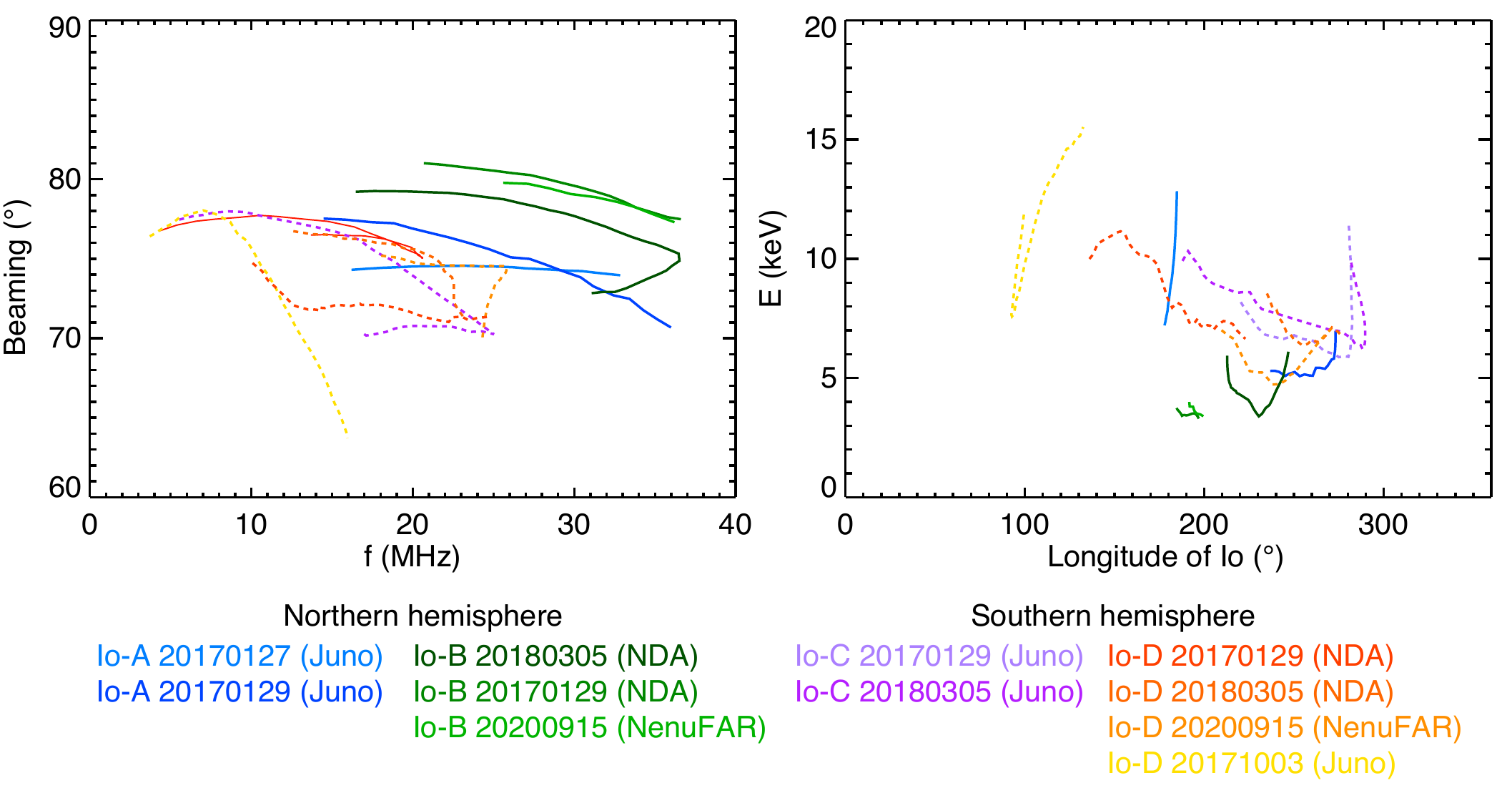}
\caption{(a) Emission angle $\theta(f)$ and (b) inferred electron kinetic energy $E_{e-}$ for 11 Io-DAM arcs observed by Juno/Waves, the NDA and NenuFAR for the dates indicated at the bottom. Having assessed the typical uncertainty on $\theta(f)$ and $E_{e-}$ resulting from the different lead angle models throughout the article, we chose here to show the results obtained from the H19 model only, for the sake of clarity. Parameters for northern (southern) emissions are plotted with solid (dashed) lines.}
\label{fig11}
\end{figure*}

The interesting method developed by \citeA{Wang_EPP_20} also relies on multi-point radio observations to accurately locate the sources. Unfortunately, the authors applied it to a single of Io-DAM event observed by Wind and STEREO A/B waves instruments, erroneously identified as an Io-B emission instead of an Io-D one \cite{Lamy_EPP_22}. The correction of the hemisphere of origin in turn yields larger $\theta(f)\sim72-65^\circ$ and lower $E_{e-}\sim5-9$~keV \cite{Wang_EPP_22}, in excellent agreement with our results summarized in Figure \ref{fig11}. The correction brought by \cite{Wang_EPP_22} also illustrated that a misidentification of the Io-DAM hemisphere yielded in their case a systematic error up to $-7^\circ$ on $\theta$ and $+9$~keV on $E_{e-}$.

\citeA{Martos_JGR_20} developed a different method, applied to one example of Io-A, B, C and D events (sometimes made of multiple arcs) as observed by Juno/Waves alone. To our understanding, their method combines fits of the position of the active IFT determined by an ionospheric lead angle and $\theta$ for each Io-DAM arc, providing ranges for this pair of covariant quantities, thus large uncertainties in a blind unconstrained fit. Among their results, the Io-D arc measured on 3rd Oct. 2017 yielded the most surprizing results, with an ionospheric lead angle as large as $40^\circ$ and low $\theta(f)$ varying between $33^\circ$ and $60^\circ$ yielding $E_{e-}\sim50$~keV. By applying our method to the same Io-D arc, we obtained the yellow dashed curves displayed on figure \ref{fig11} and the parameters listed in table \ref{tab}. These are in fair agreement with the results obtained for the 10 other Io-DAM arcs dealt with in this article, with $\theta(f)$ decreasing from $78^\circ$ to $63^\circ$, corresponding to $E_{e-}$ evolving between 7 and 16~keV. Importantly, our lead angle models placed the active IFT at $\delta\le3^\circ$ (see Figure \ref{fig1}). Our results thus significantly disagree with those of \cite{Martos_JGR_20}. Also, in their article, the authors stated that "{\it In any case, we produce better fits between our models and the observations than other recent studies \cite{Louis_GRL_17} wherein the authors obtained models that yield arcs around 1-2 hr removed from the observation times as well as MHz of difference in frequency}". It is worth adding two comments here. Firstly, the study of \citeA{Martos_JGR_20} is based on the fit of Io-DAM arcs, so that a good correspondence between the fit and the observed arc is a prerequisite of their analysis rather than a consequence of it. Secondly, the study of \citeA{Louis_GRL_17} was based on the {\it a priori} modeling of Io-DAM arcs with the ExPRES code using constant electron energies, namely 0.64keV in the north and 3 keV in the south and a simple sinusoidal lead angle model, following \citeA{Hess_GRL_08}. The purpose of \citeA{Louis_GRL_17} was therefore not to provide best fits of the observed arcs but, instead, a guide model from which any observed Io-DAM emission can be easily identified. For instance, the Io-A main arc observed on 27th Jan. 2017 in Figure \ref{fig2} occurred $\sim$1.25~h before the ExPRES-modeled arc. We refer the reader interested into the evolution of the shape of the Io-DAM arc as a function of all the parameters involved in ExPRES to the detailed parametric study of \cite{Louis_PRE8_17}. 

\begin{table*}
\center
\rotatebox{90}{
\begin{tabular}{c|c|c|c|c|c|c}
  Observer&Component&Observing time @ Jupiter&f (MHz) & Active IFT determination & $\theta(^\circ)$ & $E_{e-}$ (keV) range/median \\
  \hline
  Juno &  Io-A & 2017-01-27 12:14-12:30 & 16$-$33 & UV auroral imaging (HST) & $74-75$ & $7-13$ / $9.3\pm1.8$\\
  \hline
  NDA &  Io-B & 2018-03-05 02:15-03:29 & $16-37$ & H19 model + bi-point radio & $81-74$ & $2-6$ / $3.5\pm0.6$ \\
  NDA &  Io-D & 2018-03-05 03:03-04:32 & $12-24$ & observations ($-2.5^\circ$ correction & $78-72$ & $5-7$ / $5.6\pm0.5$ \\
  Juno &  Io-C & 2018-03-05 01:22-05:02 & $5-25$ & of the southern footprint longitude) & $77-68$ & $7-13$ / $9.8\pm1.6$ \\
  \hline
  \hline
  NDA &  Io-B & 2017-01-29 03:20-03:48 & $20-37$ & H19 model  + bi-point radio & $81-77$ & $3-4$ / $3.5\pm0.1$\\
  NDA &  Io-D & 2017-01-29 01:36-04:44 & $10-25$ &  observations (no correction & $75-71$ & $6-12$ / $8.0\pm1.6$ \\
  Juno &  Io-A & 2017-01-29 05:13-06:32 & $14-36$ & needed) & $78-70$ & $5-7$ / $5.3\pm0.53$ \\
  Juno &  Io-C & 2017-01-29 04:37-06:52 & $4-21$ & & $78-75$ & $5-12$ / $6.8\pm1.5$\\
  \hline
  NenuFAR &  Io-B & 2020-09-15 15:13-15:30 & $25-37$ & H19 model & $80-77$ & $3-4$ / $3.6\pm0.2$\\
  NenuFAR &  Io-D & 2020-09-15 15:51-18:03 & $18-26$ &  & $76-70$ & $4-8$ / $5.5\pm0.8$\\
  \hline
  Juno &  Io-D & 2017-10-03 04:19-05:46 & $3-16$ & H19 model & $78-63$ & $7-16$ / $11.4\pm2.6$\\

\end{tabular}}
\caption{Summary of the Io-DAM events (main arc) analyzed in this article and their characteristics.}
\label{tab}
\end{table*}

\section{Discussion}
\label{discussion}

\subsection{A time-frequency variable emission cone}

The results presented in the previous sections used different methods to accurately determine the Io-DAM emission angle, reaching a typical uncertainty less than a few degrees. As summarized in Figure \ref{fig11}a and in table \ref{tab}, the values of $\theta(f)$ derived for 11 arcs belonging to the A/B (north) and C/D (south) categories, all show a consistent, similar trend, with $\theta(f)$ decreasing from $75^\circ-80^\circ$ near 10~MHz to $70^\circ-75^\circ$ at maximal frequencies, with larger values in the northern hemisphere, together with a significant variability whatever the frequency, up to $10-14^\circ$. The observed decrease in frequency generally differ from that predicted by equation \ref{eq1} with a constant electron velocity. We attribute this discrepancy to additional variations of the emission cone both as a function of altitude along the active IFT and as a function of time for different longitudes of the active IFT. 

This conclusion relies on the assumption that the measured emission angle does not significantly vary as a function of the azimuthal angle around the magnetic field vector. Indeed, we neglected the flattening of the emission cone proposed by \citeA{Galopeau_JGR_16} to result from wave refraction near the source along directions probing increasing magnetic field magnitude. We briefly mentioned in section \ref{method} some limitations of the method used by \citeA{Galopeau_JGR_16} to achieve this result, but we further discuss hereafter possible implications of a strongly asymmetric emission cone on our analysis. First of all, ExPRES simulations of Io-DAM emissions have shown that the Io-B/D (A/C) observed arcs, which typically last for less than a few hours, correspond to the starting (ending) edges of the northern/southern modeled emission, whose total duration is $\sim$21~h \cite{Louis_GRL_17}. The central portion of the modeled emission is never observed, possibly as the result of a flattened emission cone with the lowest values of $\theta$ being reached for waves radiated toward the the magnetic equator, for azimuthal angles near $0^\circ$. Instead, the Io-DAM arcs are observed for limited time intervals, when Io lies near $\sim90-110^\circ$ and $\sim230-250^\circ$ phase. This configuration corresponds to waves radiated along azimuthal angles closer to $\pm90^\circ$, corresponding to raypath for which refraction near the source should play a more modest role, if any. In addition, such refraction effects should imply a systematic behaviour with a gradual decrease of $\theta$ as a function of decreasing azimuthal angles, that is as a function of increasing frequency along the arc. The effect of a flattened cone should thus add to the decrease of $\theta(f)$ expected from equation \ref{eq1}. For the Io-A emission observed on 27th Jan. 2017, the emission angle of the main Io arc (dark blue crosses in Figure \ref{fig5}) remained fairly constant with frequency, and gradually shifted from the trend predicted by equation \ref{eq1} for constant electron velocities (gray lines). More generally, we checked that $\theta(f)$ decreased less rapidly than expected from equation \ref{eq1} for constant electron velocities for 10 arcs over the 11 investigated in this study. As a result, the azimuthal variation of the emission cone was therefore fully negligible with respect to variations of $\theta$ as a function of altitude and time. A quantitative test of the flattened cone model is beyond the scope of this paper. Such a study would benefit from an update of the results of \cite{Galopeau_JGR_16} using relevant input parameters such as the JRM09 model and quantitative predictions of $\theta$ (referenced to {\bf B}) as a function of frequency and longitude.

\subsection{Variable CMI emission conditions and Io-Jupiter interaction}

The electron kinetic energies inferred from our measurements of $\theta(f)$ also vary as a function of source altitude and time/longitude. Figure \ref{fig11}b superimposes the obtained $E_{e-}$ as a function of the longitude of Io. The last column of table \ref{tab} provides in parallel the range, the median value and the standard deviation computed for each Io-DAM event. Overall, $E_{e-}$ ranges from 3 to 16~keV and displays a persisting variability, rising up to 7~keV for a given longitude of Io, much larger than the $\sim$$1-2$~keV standard deviation affecting each Io-DAM emission alone. Finally, the overall average (median) $E_{e-}$ computed from the 11 median values listed in table \ref{tab} reaches $6.6\pm2.7$~keV ($5.6\pm2.7$~keV). These results are in excellent agreement with the average value of 4~keV inferred from the study of the spectral drift of Io-DAM msec bursts (which are embedded within the long-lasting arcs investigated in this study) \cite{Hess_PSS_07} and with the 1-26~keV range derived from Juno in situ electron measurements close to the active IFT \cite{Louis_AGU_20}.

The variation of $E_{e-}$ as a function of altitude was evidenced from the curved shape of two Io-C arcs, with larger $E_{e-}$ at lower frequencies, $i.e$ at higher altitudes. A variation of the electron kinetic energy driving Io-DAM as a function of altitude is not a surprise in itself. Field-aligned potential drops, typically ranging from a few 100~eV to $\sim$1~keV have for instance been inferred from abrupts changes of the drift rate of Io-DAM msec bursts \cite{Hess_PSS_07,Hess_PSS_09}. Nevertheless, these studies showed larger energies closer to the planet (accelerated electrons propagating upward being slowed down by such potential drops), as opposed to our case studies. Applying our method to a larger sample of curved-enough arcs observed by a single observer or to emissions simultaneously observed at different frequencies by multiple observers is thus necessary to statistically assess the altitudinal evolution of $E_{e-}$ along the active IFT and to compare it to predictions of acceleration models. 

Figure \ref{fig9} showed a smooth evolution of $E_{e-}$ as a function of Io's longitude for both hemispheres, the southern electrons reaching larger velocities than the northern ones while Io was travelling in the northern part of the torus, in contradiction with the trend early derived by \citeA{Hess_PSS_10}. The variability observed in Figure \ref{fig11}b shows a more complex dynamics, likely including a large variation of $E_{e-}$ with altitude for the arcs not dealt with in Figure \ref{fig9}. It does not show a systematic behaviour per hemisphere nor clues of any simple sinusoidal variation related to Io's magnetic latitude. Here again, a statistical study over an extended number of arcs is needed to unravel any average trend and check predictions for the evolution of the Io-Jupiter Alv\'enic current with time \cite{Hess_JGR_11}.
 

Finally, by using images of the Io UV footprint or multi-point radio observations of Io-DAM arcs to constrain the position of the active IFT, we showed that the effective equatorial lead angle can shift from that predicted by our lead angle models, up to a few degrees. As two of the lead angle models are based on average UV data, this shift certainly results from an additional variation related to the plasma torus density. More precisely, the H19 lead angle model relies on a model of Alfv\'en wave propagation in the Io plasma torus in which the Alfv\'en travel time is proportional to the inverse square of the plasma density \cite{Hinton_GRL_19}. It is thus in principle possible to probe the effective Io plasma torus density from the shift between the observed and predicted equatorial lead angle. Such a diagnostic method needs to be quantitatively validated, for instance during known episodes of enhanced volcanic activity of Io such as in early 2015 \cite{Tsuchiya_JGR_18}, which is beyond the scope of this paper.

\section{Summary and perspectives}
\label{summary}

In this article, we investigated the emission angle $\theta$ of 11 cases of Io-decametric emissions belonging to the A/B (north) and C/D (south) categories, as observed by Juno/Waves, the NDA and NenuFAR. To minimize uncertainties on the determination of $\theta$, we used the up-to-date magnetic field model JRM09+C20 and developed three different methods to accurately position the active IFT hosting the radiosources producing the main arc of each Io-DAM event. These methods were respectively based on (i) updated models of the Io equatorial lead angle $\delta$, (ii) UV images of Jupiter's aurorae simultaneous to radio observations and (iii) multi-point radio measurements. We then used the measured emission angles in the loss cone-driven CMI framework to derive the kinetic energy $E_{e-}$ of source electrons accelerated by the Io-Jupiter interaction along the active IFT. 

Our results summarize as follows :

\begin{itemize}
\item Along method (i), we built up three models of $\delta$, providing the position of the active IFT for both hemispheres, based on UV average measurements of the Io footprint (B09 and B17) and on an Alfv\'en wave propagation model (H19). While still differing by a few degrees, all the models show consistent trends, with $\delta$ varying within the range $\sim$$0-10^\circ$ in the north and $\sim$$1-7.5^\circ$ in the south. They thus form a robust reference, significantly updating pre-Juno models, to predict the active IFT at any time. The comparison between the three models additionally provides a typical uncertainty on the position of the active IFT.

\item We showed that methods (ii) and (iii) are efficient means to accurately position the active IFT. For the two cases studies testing each method, the real ionospheric footprint of the active IFT was leading that predicted by method (i) by up to $2.5^\circ$, likely resulting from a time-variable velocity of Alfv\'en waves propagating within the Io plasma torus. The comparison of results provided by methods (ii) and (iii) to those obtained by method (i) may in turn form a powerful, remote, probe of the Io torus plasma density.

\item Method (ii) was illustrated with HST UV images sampling the northern Io UV footprint (with two spots dominating a series of fainter ones) while an Io-A emission (made of two bright arcs among a series of fainter ones) were simultaneously observed by Juno/Waves. The analog morphology and structure of the radio/UV emissions provided evidence that the main Io-A arc is associated with the MAW UV footprint and, for the first time, that the secondary Io-A arc is associated with the TEB UV footprint. A closer correspondence exists between radio and UV sub-structures, but was left for future investigations.

\item Multi-point simultaneous radio observations of Io-DAM, used in method (iii), can sample the waves radiated by the same radiosource along different directions and probe wide spectral (altitudinal) and temporal ranges, sometimes for both hemispheres at the same time. Multi-point radio observations thus provide a rich diagnostic of Io-Jupiter emissions all along the active IFT.

\item Overall, the measured values of $\theta(f)$ all show a consistent, similar trend, with $\theta(f)$ decreasing from $75^\circ-80^\circ$ below 10~MHz to $70-^\circ75^\circ$ at maximal frequencies, with larger values in the northern hemisphere, and significantly vary both as a function of frequency (altitude) and time (Io's longitude). The variation of the emission cone as a function of altitude can be partly explained by the expected theoretical decrease of $\theta$ with increasing magnetic field, that is with increasing frequency. The uncertainty on $\theta$ is dominated by the uncertainty on the longitude of the active IFT, and typically reaches a few degrees.

\item The inferred $E_{e-}$, assuming the validity of equation \ref{eq1}, lie between 3 and 16~keV, reaching an average (median) value of $6.6\pm2.7$~keV ($5.6\pm2.7$~keV). $E_{e-}$ also varies, by a few keV, both as a function of frequency (altitude) and time (Io's longitude). The former variation was generally not taken into account in past studies of the Io-DAM emission angle but plays a role as significant as the variation with time. The observed variations do not compare with those published in the literature and need to be statistically investigated to deconvolve each source of variation. 
\end{itemize}

A natural perspective of this proof-of-concept study is to take advantage of long-term radio observations of Jupiter acquired by either space-based observatories or ground-based radiotelescopes to increase the statistics of Io-DAM events and perform a statistical analysis of $\theta$ and $E_{e-}$, and establish their average altitudinal and temporal variations. The knowledge of the altitudinal profile of $E_{e-}$ could for instance help to constrain electron acceleration models along the active IFT, while measuring the evolution $E_{e-}$ as a function of Io's longitude or as a function of time on long-term scales provides a remote probe of the Io-Jupiter interaction. A statistical investigation is also needed to quantify any geometrical asymmetry of the emission cone, expected from wave refraction near the source, which we showed to be negligible here. Interestingly, several radio observatories already accumulated Jupiter observations through years, providing an ideal dataset to conduct a statistical study. The NDA quasi-daily observes Jupiter over $10-40$~MHz since January 1978, and various catalogs of its Routine dataset, such as that of \cite{Marques_AA_17} which extend from 1990 up to now, make it straightforward to identify Io-DAM events. Since 1993, the Waves instrument onboard the Wind spacecraft also continuously records radio emissions from Jupiter up to 13~MHz. Finally, the continuous radio observations of Juno/Waves (0-40~MHz) since 2016, of Cassini/RPWS (0-16~MHz) in 2000-2001 and of Voyager/PRA (0-40~MHz) in 1978-1979 provided close-in measurements covering a long period of time.

\section{Open research}
The Juno/Waves data used in this article are publicly accessible through the NASA Planetary Data System at \url{https://pds.nasa.gov}. The NDA/Routine public dataset can be accessed from the Nan\c cay portal at \url{https://obs-nancay.fr} \cite{Cecconi_NDA-Routine_20}. The HST/STIS data were retrieved from the public APIS service at \url{https://apis.obspm.fr} \cite{apis_hst_data}. The NenuFAR/UnDySPuTeD early science observation have been referenced for the purpose of this study as \cite{nenufar_jupiter_data}.

\acknowledgments
The authors thank the Juno, NDA, NenuFAR, HST/GO 14634 and APIS teams for the data acquisition, processing and release to the community. We acknowledge the use of the Nan\c cay Data Center computing facility (CDN - Centre de Donn\'es de Nan\c cay). The NDA, NenuFAR and CDN are hosted by the Station de Radioastronomie de Nan\c cay in partnership with Observatoire de Paris, Universit\'e d'Orl\'eans, OSUC, R\'egion Centre Val de Loire and the CNRS. The APIS service is operated at LESIA/Paris Astronomical Data Centre (PADC) with the support of Observatoire de Paris. The Ju

LL thanks in particular S. Aicardi for maintaining the Jupiter probability tool, D. Afgoun for her work on the centering of Jupiter's HST images during her M1 internship, S. Hess and B. Bonfond for useful discussions on Io-driven processes, R. Desmonts for having built the catalog of HST images of Jupiter used in this article during a research discovery week, P. Hilton and F. Bagenal for having provided the footprint coordinates of their model, L. Denis and A. Coffre for having made the NDA working almost every day all these years. The authors were supported by CNES and CNRS/INSU programs of planetology and heliophysics.

\end{document}